\documentclass[aps,pre,preprint,groupedaddress]{revtex4}

\usepackage{graphicx}
\usepackage{url}

\def\gappeq{\mathrel{ \rlap{\raise.5ex\hbox{$>$}}
                      {\lower.5ex\hbox{$\sim$}}  } }
\def\lappeq{\mathrel{ \rlap{\raise.5ex\hbox{$<$}}
                      {\lower.5ex\hbox{$\sim$}}  } }

\begin{document}

\title{Formation of Multifractal Population Patterns from Reproductive Growth and Local Resettlement}

\author{Jonathan Ozik}\email{jozik@umd.edu}
\affiliation{Department of Physics and Institute for Research in Electronics and Applied Physics, University of Maryland, College Park, Maryland 20742}

\author{Brian R. Hunt}
\affiliation{Department of Mathematics and Institute for Physical Science and Technology, University of Maryland, College Park, Maryland 20742}

\author{Edward Ott}
\affiliation{Department of Physics, Department of Electrical and Computer Engineering, and Institute for Research in Electronics and Applied Physics, University of Maryland, College Park, Maryland 20742}

\date{\today}

\begin{abstract}

We consider the general character of the spatial distribution of a population that grows through reproduction and subsequent local resettlement of new population members. We present several simple one and two-dimensional point placement models to illustrate possible generic behavior of these distributions. We show, numerically and analytically, that these models all lead to multifractal spatial distributions of population. Additionally, we make qualitative links between our models and the example of the Earth at Night image~\cite{ea}, showing the Earth's nighttime man-made light as seen from space. The Earth at Night data suffer from saturation of the sensing photodetectors at high brightness (`clipping'), and we account for how this influences the determined dimension spectrum of the light intensity distribution.

\end{abstract}


\maketitle

\section{Introduction}

\begin{figure*}
\includegraphics[width=6.25in]{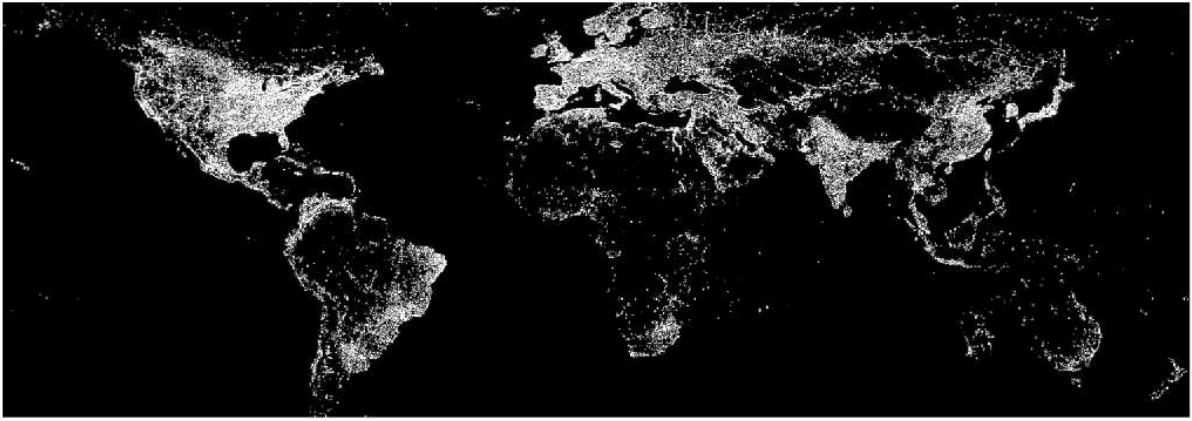}
\caption{\label{eatn}A version of the Earth at Night image~\cite{ea}. This is a composite image created with data from the Defense Meteorological Satellite Program of the Earth's nighttime man-made lights as seen from space. For the analysis, a grayscale 30000x15000 pixel image was used. The image pixels have intensity values ranging from 0 to 100.}
\end{figure*}

Growing populations exist in many different areas of interest. Although the most obvious examples are biological in nature (e.g., bacterial cultures, the human population), one can also consider the growth of technological development or the growth of urban infrastructure as more abstract examples.
While the specific details of these systems can be quite different, the examples stated here all share the characteristic that their populations grow on some background space. A reasonable question to ask then is what types of spatial distributions result in these growing systems.
Figure~\ref{eatn} is a version of the popular Earth at Night image (EaN)~\cite{ea}. This is a composite image taken by orbiting satellites that shows the Earth's nighttime man-made lights as seen from space.
The light intensities are brightest in areas that are known to be highly populated and developed, and thus the image can be crudely thought of as representing the spatial distribution of some combined measure of technological development and human population density. We see a very heterogeneous distribution, with areas of very high intensities as well as areas of almost no light at all.

A useful characterization of the heterogeneity of a distribution 
is the distribution's {\it dimension spectrum}  $D_q$. A grid of hypercubes, or boxes, of a fixed size with edge length $\epsilon$ is used to cover the distribution, where the dimension of the hypercubes corresponds to the dimension of the underlying space (e.g., squares  for two dimensions and line segments for one dimension). $D_q$ is then calculated via~\cite{proc,grass1}
\begin{equation}
\label{dq}
D_q={1 \over 1-q}\lim_{\epsilon \to 0}{\ln I(q,\epsilon) \over \ln(1/\epsilon)},
\end{equation}
where
\begin{equation}
\label{Iq}
I(q,\epsilon)=\sum^{N(\epsilon)}_{i=1}\mu^{q}_{i}.
\end{equation}
Here the sum is taken over $N(\epsilon)$ non-empty grid boxes and $\mu_{i}$ is the fraction of the particular quantity of interest (assumed to be nonnegative, e.g., light intensity in Fig.~\ref{eatn}) that is contained in box $i$.

The parameter $q$ ($q\geq0$ in this paper) can be varied continuously to vary the influence of high and low $\mu_i$ boxes, with larger values of $q$ emphasizing the boxes with larger $\mu_i$'s. 
In particular, when $q=1$ we have the {\it information dimension} $D_1$ which, applying L'Hospital's Rule to (\ref{dq}) and (\ref{Iq}), is given by
\begin{equation}
\label{d1}
D_1 = \lim_{\epsilon\to 0} {\sum^{N(\epsilon)}_{i=1}\mu_i \ln \mu_i \over \ln \epsilon}.
\end{equation}
A distribution is {\it fractal} if it possesses a fractional (non-integer) dimension (i.e., if $D_q$ is not an integer for some $q$). If $D_q$ depends on $q$, then the distribution is said to be {\it multifractal}.

\begin{figure}
\includegraphics[width=5in]{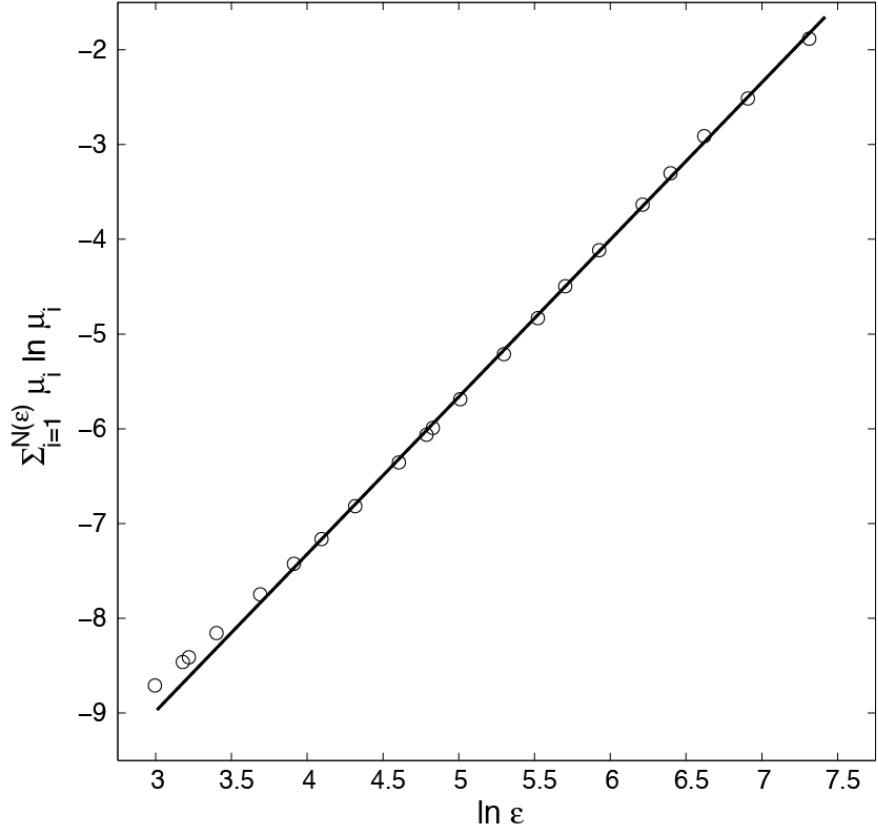}
\caption{\label{eatd1}Plot for calculating the information dimension $D_1$ [see Eq.~(\ref{d1})] for North America in the Earth at Night image. The quantity $\sum^{N(\epsilon)}_{i=1}\mu_i \ln \mu_i$ is plotted versus $\ln \epsilon$ (open circles), where $\epsilon$ is in units of pixels, resulting in a linear {\it scaling region}, the slope of which is a finite-scale approximation of $D_1$. The solid line has a slope of $1.65 \pm 0.02$, where the error value indicates the uncertainty involved in extracting the slope of the scaling region.}
\end{figure}

Generally, when measuring the dimension of a distribution of numerical and/or experimental data, there is a maximum resolution observable. In the EaN image, for example, the image's individual pixels are the smallest boxes that can be used to cover the distribution in the calculation of $D_q$. In such cases the $\epsilon\to0$ limit cannot be considered; instead one looks for a {\it scaling range} of $\epsilon$ where the quantity $(1-q)^{-1}\ln I(q,\epsilon)$ varies approximately linearly with $\ln (1/\epsilon)$, from which a dimension can be extracted as the slope of a straight line fitted to data in the $\epsilon$ scaling range. We illustrate this process by calculating  the information dimension $D_1$ of North America (6000x3000 pixels) in the EaN image.
Figure~\ref{eatd1} shows a plot of $\sum^{N(\epsilon)}_{i=1}\mu_i \ln \mu_i$ versus $\ln \epsilon$  (with $\epsilon$ in units of pixels) for the image, where each $\mu_i$ is the fraction of the total intensity in box $i$. The good evidence of linearity in this plot over a scaling range in $\epsilon$ of order $e^4\approx 50$ indicates that a fractal description makes sense. The solid line in Fig.~\ref{eatd1} has a slope of $1.65 \pm 0.02$, which we take as the value of $D_1$ (the error indicates the uncertainty involved in extracting the slope of the scaling region). Thus the light distribution over North America in the EaN is fractal. (Other regions of the EaN also exhibit fractality, with different values of $D_1$.) In Sec.~\ref{clip} we demonstrate that the EaN light distribution is also multifractal.

Yook et al.~\cite{barabasi} have recently reported that the world's human population, as well as the population of internet routers and autonomous systems, are fractally distributed. Additionally, several  studies~\cite{batlon,makhavsta,app1,manzansol,samvos,app2} have demonstrated various fractal properties of urban settlements. We were thus led to consider the possibility of common, underlying mechanisms shared by  growing populations that lead to fractal distributions.

One characteristic found in many growing populations is the existence of some generalized reproduction process, where existing members of the population generate more members. In addition to biological reproduction, in some non-biological systems, concentrated populations can encourage the creation of more members. For example, areas that are technologically highly developed are likely to stimulate more development than areas that are not as technologically developed.

Generally in such systems, new members of a population become situated near an existing member, but not at the exact same location. We model this process as a local ``resettlement", whereby an existing member of the population produces one or more new members, who then resettle to a nearby location.

In this paper we demonstrate that these two ingredients, reproduction and local resettlement, can lead to multifractal spatial distributions for growing populations. We present several point placement models in one and two dimensions which implement these ideas and show, both numerically and analytically, that they lead to multifractal distributions.

One might suspect that fractal population densities arise due to strong inhomogeneities of the underlying space (e.g., inhomogeneities in the distribution of land fertility, natural resources, etc.). Our models initially (Sec.~\ref{models}) employ spaces that have {\it no inhomogeneity}, thus demonstrating that fractal population distributions can occur without inhomogeneity of the background space on which the population grows. We also consider generalizations to our models that include geographical inhomogeneities (Sec.~\ref{inhom}) and demonstrate that, for the particular forms of the inhomogeneities that we employ, the multifractality of the distribution is unaffected.

In Sec.~\ref{models} we present our point placement models along with numerical results. (We emphasize that the set of models we introduce in Sec.~\ref{models} is not exhaustive, and many other similar models could be conceived.)
Section~\ref{clip} considers situations where a multifractal distribution is sensed by an instrument that saturates at a maximum measurable value (clipping). We show that such clipping occurs in the EaN data, affecting the determined $D_q$.
In Sec.~\ref{inhom}, we investigate the effect of adding geographical inhomogeneity to the underlying space of our models. In Sec.~\ref{theory} we present analytical results for some of our models. We summarize our findings and conclude in Sec.~\ref{conclusion}. Appendix~\ref{pf} provides details and background on the analytical calculation of the spectra of fractal dimensions for some of our models. Appendix~\ref{ec} contains the proof of a theorem used in Appendix~\ref{pf} (this theorem should be of very general utility, not restricted to our specific models, in analysis of the spectra of fractal dimensions).

\section{\label{models}Models and Results}

\subsection{\label{m1}1-D Random Interval (Model 1)}

\begin{figure}
\includegraphics[width=4in]{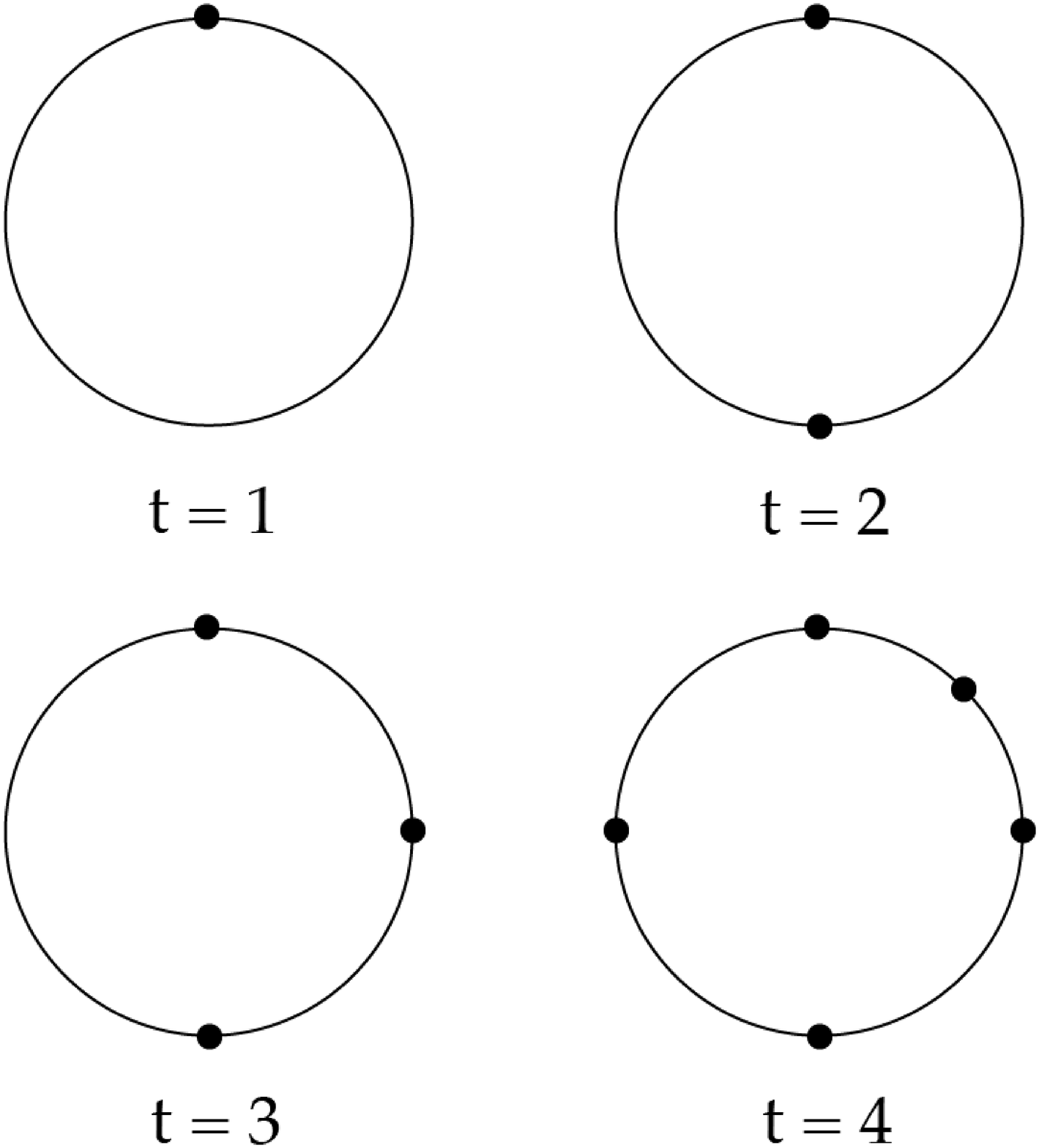}
\caption{\label{lineconst}Three representative steps of Models 1 and 3.}
\end{figure}

We begin with a very simple model that places points on the circumference of a circle. (For convenience we consider a circumference of unit length.) Our initial state is a single point on the circle's circumference. As the population (number of points on the circumference of the circle) grows, we consider the intervals of the circle's circumference between two adjacent points. At each subsequent discrete time step (each `generation') labeled $t$, a new point is `born' in the middle of each pre-existing interval with the same given fixed birth probability $p$, with each new point bisecting its corresponding target interval. Several representative steps of this construction are illustrated in Fig.~\ref{lineconst}.

In this way, every interval has an equal probability to generate or attract new points at the same rate $p$. This also implies that regions with a high density of points (i.e., with a greater number of intervals) are more likely to attract new members than areas that are sparsely populated, creating a ``rich get richer" phenomenon. We will see that all the models presented in this paper have this general characteristic. The simple rule of placing each new point in the center of an interval allows us to obtain analytical results, as will be shown in Sec.~\ref{theorym1}.

\begin{figure}
\includegraphics[width=5in]{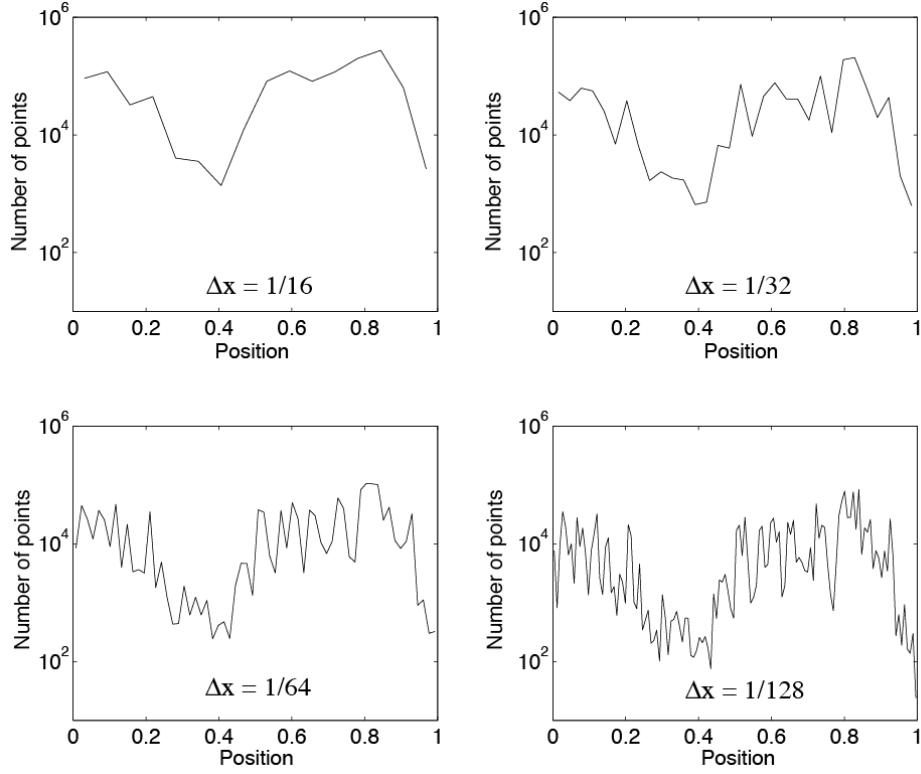}
\caption{\label{log_rho}Semilog plots of histograms of point locations on the unit length circumference of a circle for a point distribution ($\sim 10^6$ points) generated by Model 1 with $p=0.5$, at time $t=35$, for various bin sizes $\Delta x$. a) $\Delta x=1/16$, b) $\Delta x=1/32$, c) $\Delta x=1/64$, d) $\Delta x=1/128$. }
\end{figure}

For $p=1$ each existing interval is bisected by a point at each time step, resulting in a uniformly distributed set of points on the circle circumference for all $t$. Thus, for more interesting point distributions, we turn our attention to models with $p<1$. As an example, we numerically examine the $p=0.5$ case. Figure~\ref{log_rho} shows semilog plots of point location histograms, along the circle circumference, with different bin sizes $\Delta x$ for one realization of our point placement scheme with $p=0.5$, at time $t=35$. We observe a roughening of the plot as smaller histogram bins are used, thus giving a sense of the heterogeneity of the distribution. To quantify this heterogeneity, we calculate the dimension spectrum $D_q$ of the distribution. We cover the distribution of points with intervals of a fixed length $\epsilon$ and assign to each interval $i$ a measure $\mu_i$ equal to the fraction of the total points contained in $i$. Then, referring to Eq.~(\ref{dq}), we plot the quantity $(1-q)^{-1}\ln I(q,\epsilon)$ versus $\ln(1/\epsilon)$ for a range of $\epsilon$, so that the slope of the graph in the scaling range gives us $D_q$.

For $q \geq 0.5$ the plots exhibit a linear regime that allows for a relatively unambiguous determination of the dimension. However, for smaller $q$ there tend to be shorter scaling regions, which makes the extraction of a reliable $D_q$ more error prone. This is due to the fact that the distribution is made up of a finite collection of points. As the grid covering the distribution is made finer, grid intervals in the more sparsely populated areas begin to contain either 0 or 1 points. Furthermore, if $\epsilon$ is made small enough, eventually all the points in the distribution are covered by different boxes, and the dimension for the distribution is simply the dimension of a collection of individual points, i.e., zero. Thus, in order to get a meaningful dimension out of a finite distribution of points, a certain degree of coarseness of the covering grid is necessary. For small values of $q$, $D_q$ relies heavily on boxes containing small amounts of the measure. Therefore, as $\epsilon$ is decreased, a scale is quickly reached where individual points in sparsely populated areas are covered by separate boxes, thereby destroying the overall scaling behavior. For larger $q$, the dense regions have a large enough influence on $D_q$ that the effects of the finiteness of the distribution are not observed until very small scales are reached, allowing for a relatively larger scaling range.

\begin{figure}
\includegraphics[width=6.25in]{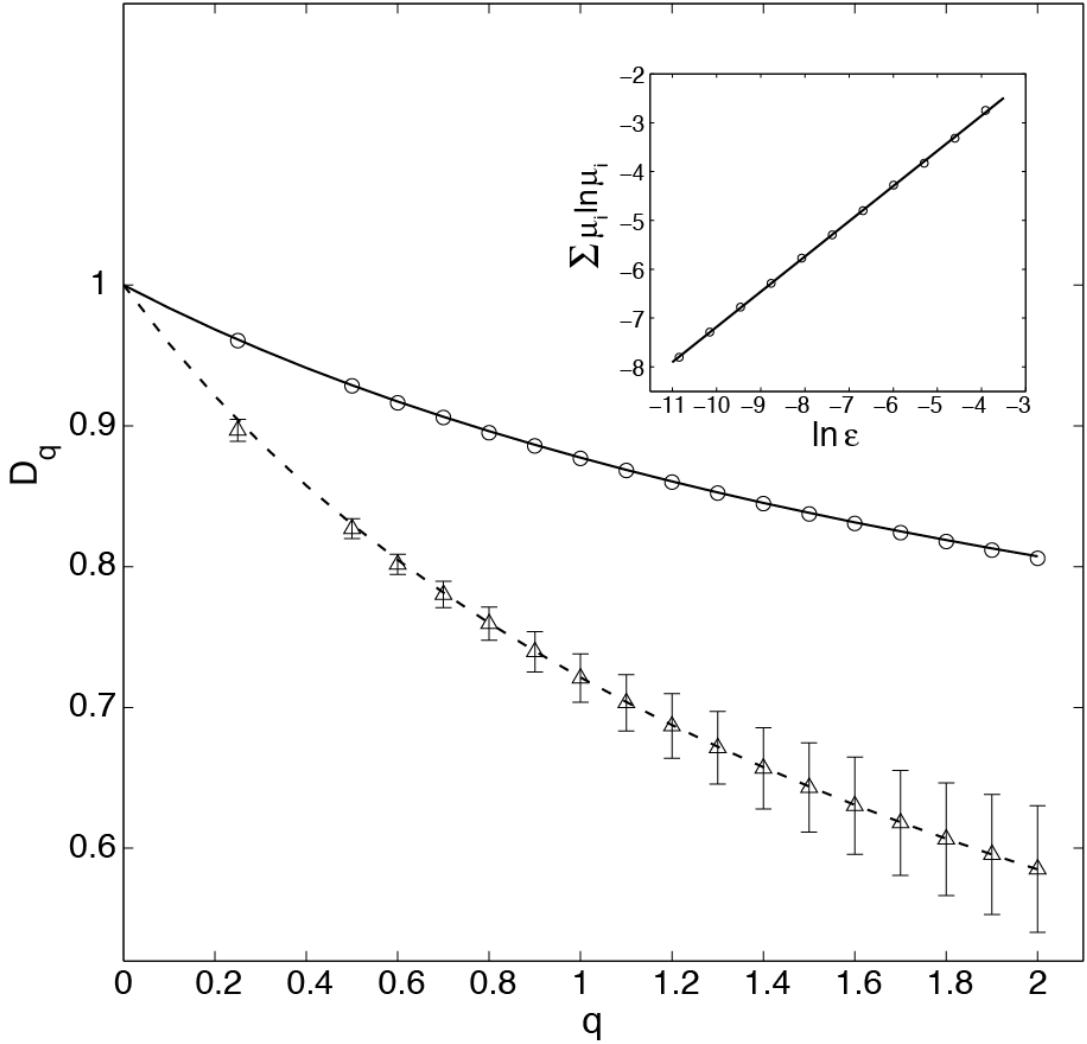}
\caption{\label{dqvsq1}Plots of $D_q$ for Model 1 with $p=0.5$ and $p\to 0$. The open circles ($p=0.5$) and triangles ($p\to 0$) are the numerical values calculated from the slope of a line fit to $(1-q)^{-1}\ln I(q,\epsilon)$ versus $\ln(1/\epsilon)$ [see Eq.~(\ref{dq})], averaged over 20 realizations of point distributions (except for $q=1$, in which case the line is fit to $\sum^{N(\epsilon)}_{i=1}\mu_i \ln \mu_i$ versus $\ln \epsilon$ [see Eq.~(\ref{d1})]), with the error bars indicating the sample standard deviation for each value (in the $p=0.5$ case the error bars are all smaller than the open circles and we omit them). The solid line is the analytical result of Eq.~(\ref{dq1}) with $p=0.5$, while the dashed line corresponds to Eq.~(\ref{dq1p0}). As an example the inset shows the determination of $D_1$ for one point distribution. For $p=0.5$, the distributions are calculated at time $t=35$ and have between $1.2 \times 10^5$ and $2.8 \times 10^6$ points. For $p\to 0$ the numerical results were obtained for distributions of $10^6$ points.}
\end{figure}

Figure~\ref{dqvsq1} is a plot of the numerically determined average $D_q$ (open circles) for 20 realizations of point distributions generated by Model 1 with $p=0.5$ at time $t=35$. (The number of points in each realization ranges from $1.2 \times 10^5$ to $2.8 \times 10^6$ points.) The solid line represents the analytical result for $D_q$ derived in Sec.~\ref{theorym1} (Eq.~(\ref{dq1}) with $p=0.5$). Both the numerical and analytical results demonstrate that this model generates multifractal distributions.

Notice that our analytical value for the box-counting dimension $D_0$ is
one.  Unlike $D_q$ for $q$ positive, $D_0$ depends only on the limiting
set of points and not their distribution.  Our result that $D_0 = 1$
indicates that, although the point {\it distribution} is multifractal, the {\it set} filled by the points in Model 1 as time $t$ goes to
infinity is not fractal.  In fact, the entire circumference is filled with
probability one, which can be easily seen as follows. Since each interval has a probability $(1-p)$ of not being bisected at each time step, the probability for an interval not to have been bisected by time $t$ is simply  $(1-p)^t$, which approaches zero as $t$ approaches infinity. Therefore no empty intervals remain as $t \to \infty$. One can similarly argue for the other models in this paper (except Model 7) that the limiting sets are not fractal; the heterogeneous point distributions they yield are reflected by the fractional values of $D_q$ for $q > 0$.

We can also consider Model 1 in the limit where $p\to 0$. We think of this limit as the situation where at most one interval is chosen to be bisected from all available intervals at each time step. This allows us to reformulate the point placement scheme as follows. We begin with the same initial condition as the original model, which is a single point on the circle circumference. Then at each subsequent discrete time step we place a new point according to the following prescription:
(a) a target point, or ``parent", is chosen from all of the preexisting points with equal probability;
(b) the new point, or ``child", is placed in the middle of one of the two empty intervals adjacent to the target point with equal probability.
Steps (a) and (b) are repeated until the desired system size is reached.

To show that this point placement scheme coincides with the $p\to 0$ limit of Model 1, we plot the numerically determined average $D_q$ for 20 realizations of $10^6$ point distributions generated by it in Figure~\ref{dqvsq1} (open triangles) and observe that the data agree with the $p\to 0$ limit of the analytical result for $D_q$ (dashed line) calculated in Sec.~\ref{theorym1} [Eq.~(\ref{dq1p0})].

Thus we see that Model 1 can create point distributions with $D_q$ ranging from the trivial $D_q=1$ curve, corresponding to $p=1$, to the $D_q$ curve associated with  $p\to 0$.

\subsection{\label{m6}2-D Square (Model 2)}

\begin{figure}
\includegraphics[width=4in]{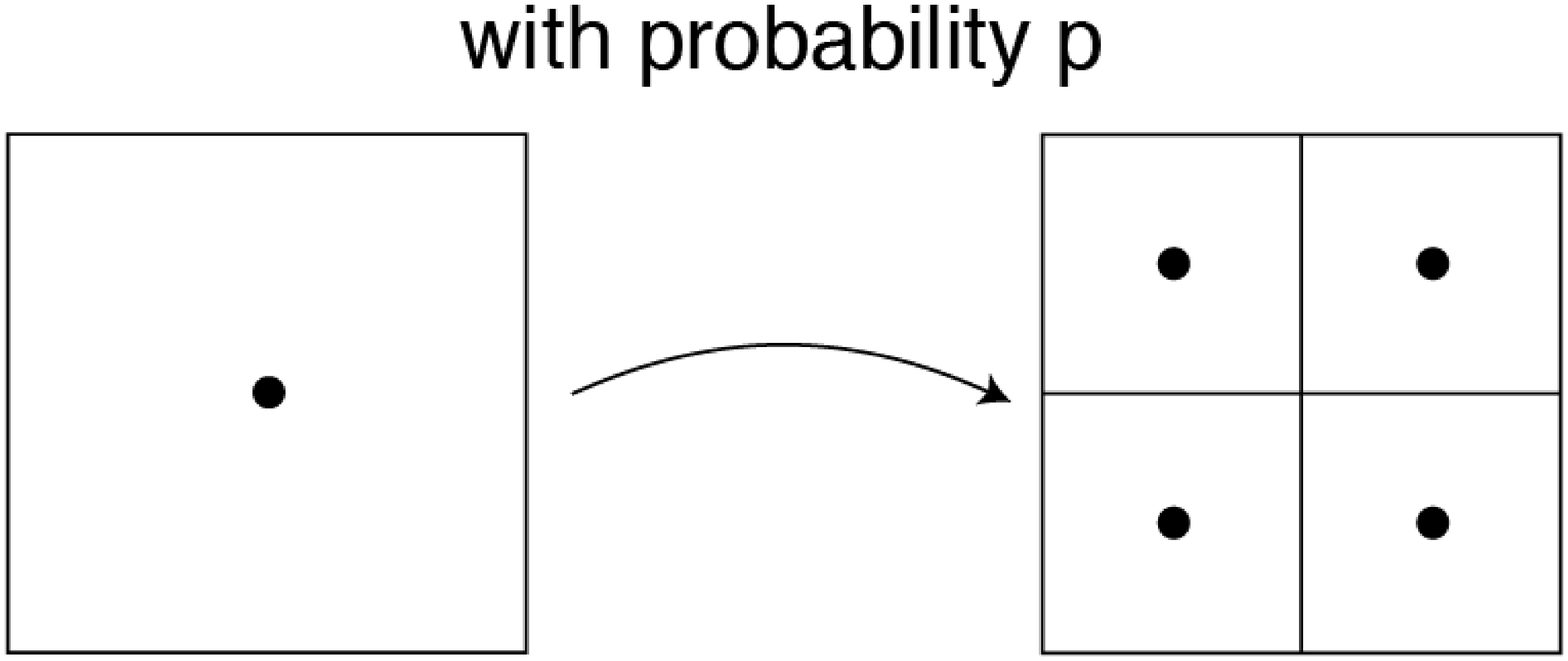}
\caption{\label{sqfig}Illustration of the four-fold ``cell division" step in Model 2.}
\end{figure}

Model 2 is a two dimensional analogue of Model 1, where points are placed on the unit square, ($0\leq x\leq1,~0\leq y\leq1$), instead of the circumference of a circle. Beginning with a single point at $(0.5, 0.5)$, we consider reproduction by a process mimicking cell division. At each discrete time step, each preexisting point, or ``parent", has a probability $p$ of dividing into four points, or ``children", where these four points disperse to occupy (resettle) the centers of the four equal size squares obtained by partitioning the original square associated with the parent point; see Fig.~\ref{sqfig}.

Note that this construction can be specified without reference to points, but only to the squares they occupy: At each discrete time step, each square is subdivided into four equal new squares with probability $p$. After the desired iterations of this area partitioning procedure, the corresponding population distribution can be produced by simply placing a point in the center of each square.

\begin{figure}
\includegraphics[width=5in]{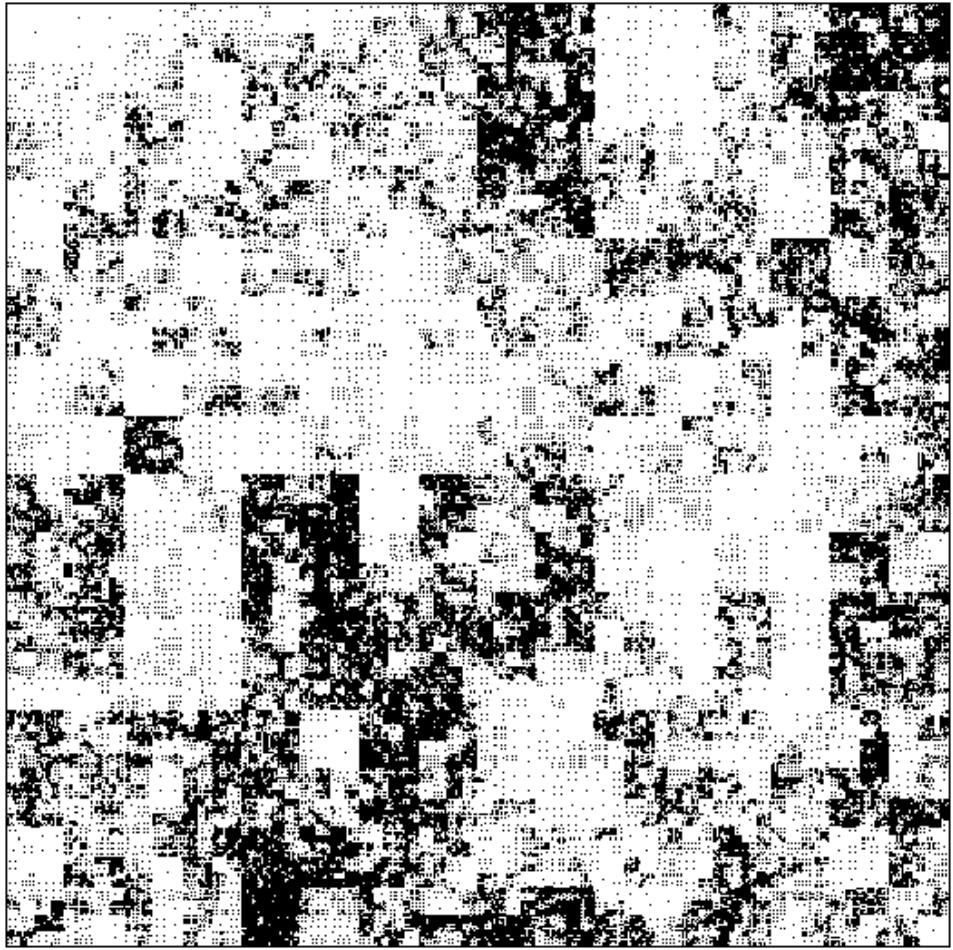}
\caption{\label{2dsq0p5}A plot of a point distribution on the unit square generated by Model 2 with $t=15$ and $p=0.5$.}
\end{figure}

\begin{figure}
\includegraphics[width=5in]{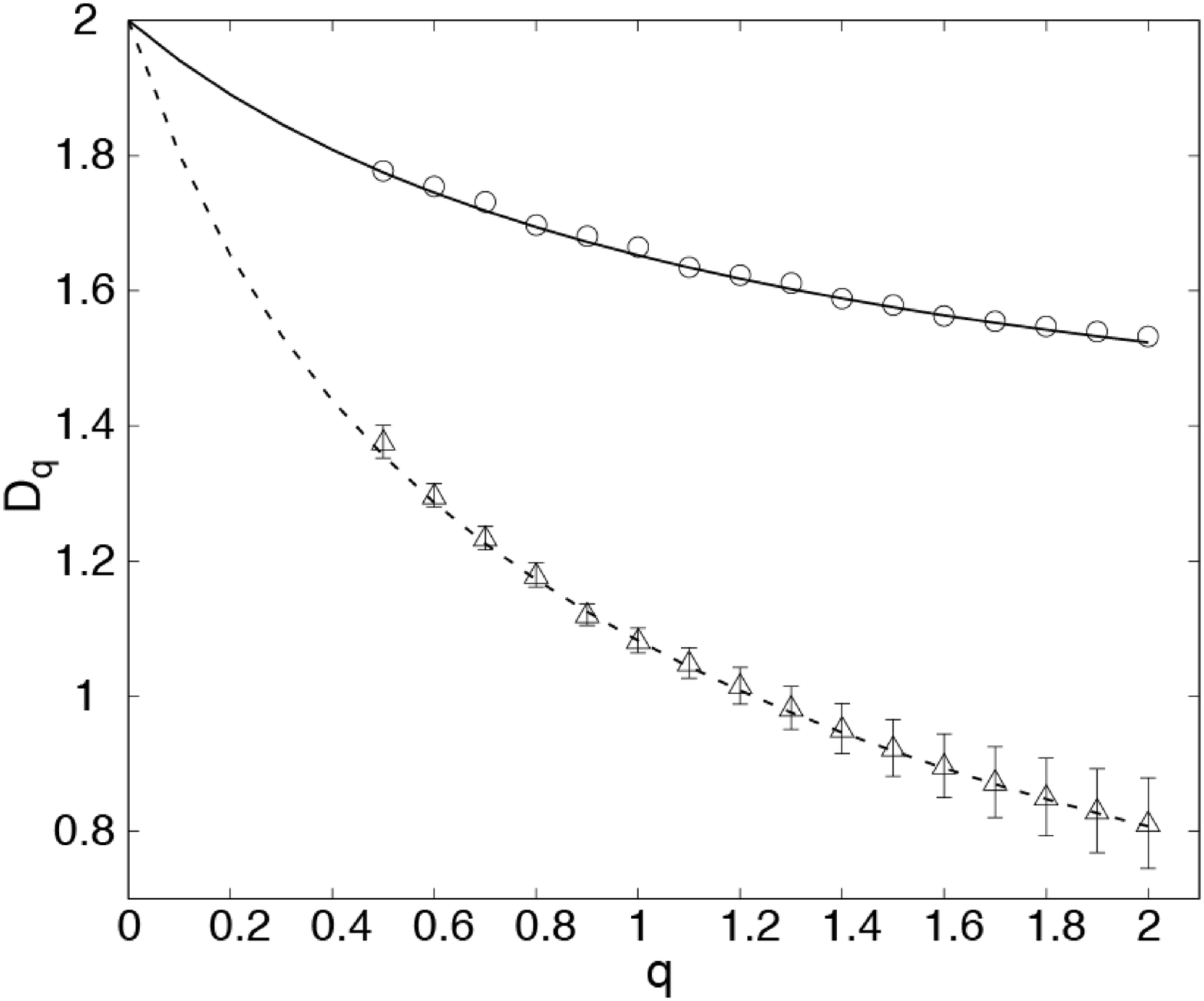}
\caption{\label{dqvsq6} Plot of $D_q$ for Model 2 with $p=0.5$ and $p\to 0$. The open circles  ($p=0.5$) and triangles ($p\to 0$) are the average numerical values of $D_q$ (see caption in Fig.~\ref{dqvsq1}) obtained from 20 realizations of point distributions generated by this model and the error bars are the sample standard deviations for each value (in the $p=0.5$ case the error bars are all comparable in size to the open circles and we omit them). The solid line is the analytical result of Eq.~(\ref{dq23}) and the dashed line is Eq.~(\ref{dq2p0}). For $p=0.5$, the distributions are calculated at time $t=15$ and have between $1.2 \times 10^5$ and $2.9 \times 10^6$ points. For $p\to 0$ the numerical results were obtained for distributions of $10^6$ points.}
\end{figure}

Since $p=1$ results in a homogeneous planar distributions of points ($D_q = 2$), we again consider $p<1$ for more interesting point distributions. Choosing $p=0.5$, we plot in Fig.~\ref{2dsq0p5} a distribution of points generated by this model at time $t=15$. The points form a heterogeneous distribution with dense clusters in some regions and relatively sparsely populated areas in other regions. Figure~\ref{dqvsq6} is a plot of the numerically determined average $D_q$ (open circles) for 20 realizations of point distributions generated by Model 2, with $t=15$ and $p=0.5$. The solid line represents the analytical result for $D_q$ derived in Sec.~\ref{theorym23}.

As we did for Model 1, we can explore the $p\to 0$ limit by considering the construction of the point distribution one point at a time. The appropriate process is as follows: At each time step a ``parent" point is chosen out of all preexisting points with equal probability and produces four ``children" points according to the four-fold cell division process described above. The open triangles in Fig.~\ref{dqvsq6} are the average $D_q$ of 20 independent $10^6$ point distributions generated by this construction, while the dashed line is the $p\to 0$ limit of the theoretical result (Sec.~\ref{theorym23}). Thus, Model 2 creates point distributions with $D_q$ ranging between $D_q=2$ for $p=1$ and the $D_q$ associated with  $p\to 0$, and hence, other than the $p=1$ case, results in multifractal point distributions.

\subsection{2-D Triangle (Model 3)}
\begin{figure}
\includegraphics[width=5in]{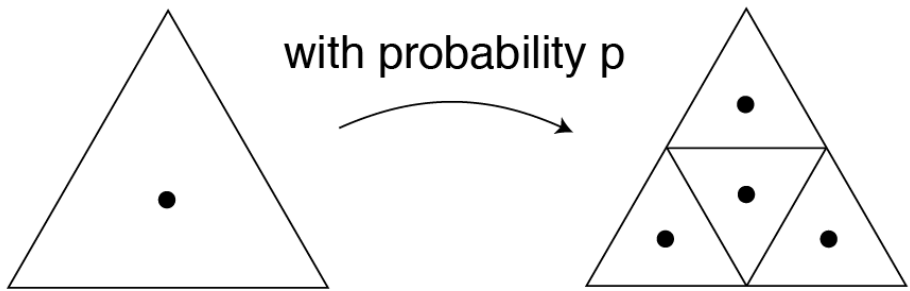}
\caption{\label{trifig}Illustration of the ``reproductive" step in Model 3.}
\end{figure}

Model 3 is a slight variation of Model 2. Here triangles, as opposed to squares, are the basis of the construction. We begin with a point in the middle of a single equilateral triangle with unit edge length. At each time step each preexisting triangle (parent) is divided into four identical equilateral triangles (children) with probability $p$, with the edge lengths of the child triangles equal to half the edge length of the original parent triangle (see Fig.~\ref{trifig}) and with points placed in the centers of the child triangles.
We show in Sec.~\ref{theorym23} that Model 3 has the identical analytical expression for $D_q$ as Model 2, and thus also results in multifractal distributions.

\subsection{2-D Random Square (Model 4)}
\begin{figure}
\includegraphics[width=4in]{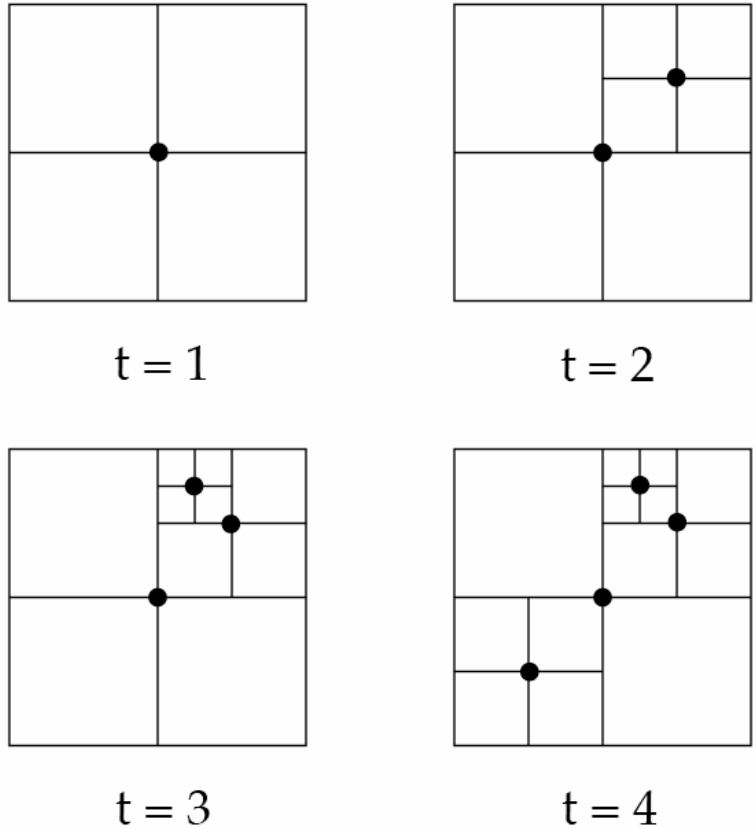}
\caption{\label{sqconst}Three representative steps of Models 4 and 6.}
\end{figure}

Model 4 is a two dimensional point placement scheme similar to the $p\to 0$ limit construction of Model 2, where points are placed on the unit square, ($0\leq x\leq1,~0\leq y\leq1$). We begin with a single point at $(0.5, 0.5)$. Imagining vertical and horizontal lines drawn through this point, we see that the point is at the vertex of four neighboring squares. At the next time step, we choose one of the squares at random with equal probability, and place a point in its middle. We then divide this chosen square into four smaller squares by horizontal and vertical lines through the newly added point. Then we continue this process:
At each subsequent discrete time step we pick a target point (parent) from all preexisting points with equal probability and add a new point (child) to the center of one of the target point's four adjacent squares, also chosen with equal probability. Several representative steps of this construction are illustrated in Fig.~\ref{sqconst}.

Notice that some squares have two potential parent points at their corners and some squares have only one. The former squares are twice as likely to be subdivided as the latter squares, unlike the $p\to 0$ limit of Model 2, in which all squares are equally likely to be subdivided.

\begin{figure*}
\includegraphics[width=6.25in]{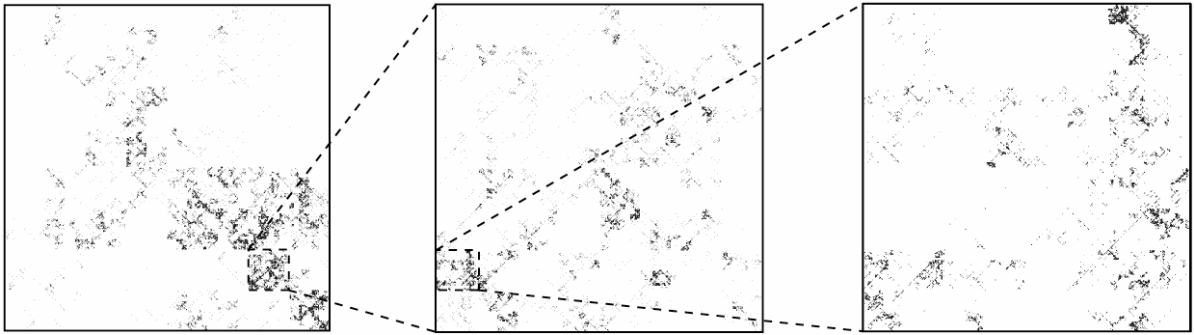}
\caption{\label{2dsq}A plot of a $4\times10^5$ point distribution on the unit square generated by Model 4. Dense regions are magnified to illustrate the heterogeneity of the point densities in the distribution at different scales.}
\end{figure*}

\begin{figure}
\includegraphics[width=5in]{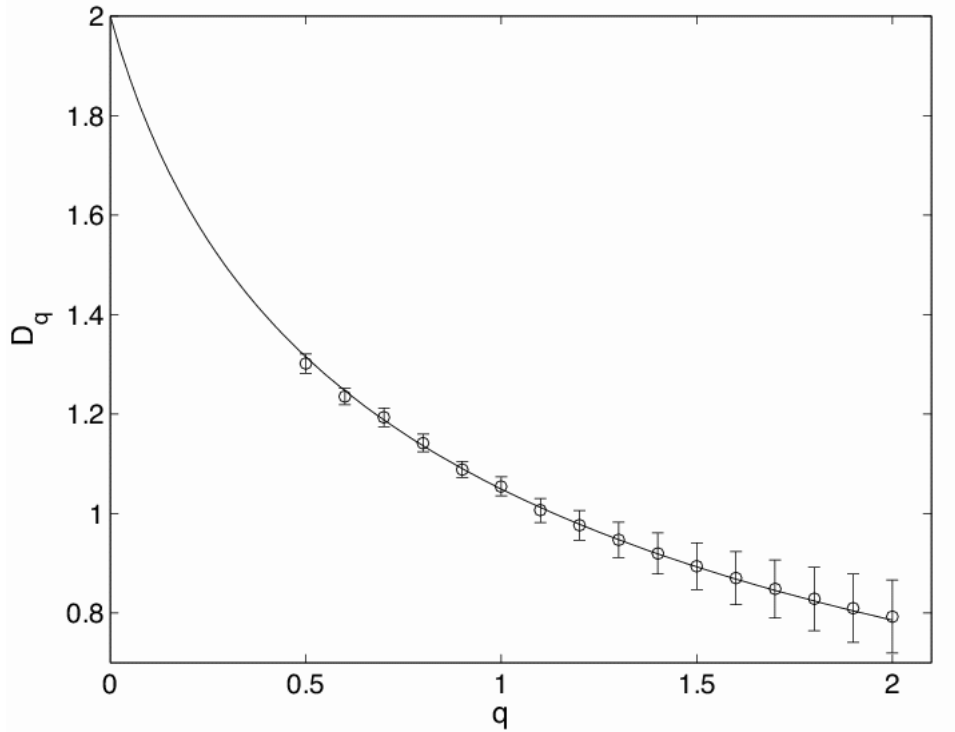}
\caption{\label{dqvsq2}Plot of $D_q$ for Model 4. The open circles are the average numerical values of $D_q$ (see caption in Fig.~\ref{dqvsq1}) obtained from 20 realizations of $10^6$ point distributions generated by this model and the error bars are the sample standard deviations for each value. The solid line is the analytical result of Eq.~(\ref{dq4}).}
\end{figure}

Figure~\ref{2dsq} shows a distribution generated by this model's point placement scheme, where we have successively magnified dense regions of the distribution to illustrate the large differences in densities that result at various length scales. Figure~\ref{dqvsq2} is a plot of the average $D_q$ of 20 independent distributions generated by this model, where the solid line is the analytical result (\ref{dq4}) derived in Sec.~\ref{theorym4}. Both the numerical and analytical results confirm that Model 4 generates multifractal distributions.

\subsection{1-D Larger Interval (Model 5)}
We return now to one dimension with a model similar to the construction of the $p\to 0$ limit of Model 1, but with a different local resettlement rule. Referring to the $p\to 0$ point placement prescription of Model 1, the initial state and Step (a) are identical while Step (b) is modified. Instead of a new point being randomly placed in one of the adjacent intervals of the chosen target point with equal probability, the new point is placed in the larger of the two intervals adjacent to the target point, and, if the two intervals are equal in size, one is chosen at random.

In this way, the new points seek out more sparsely populated local intervals to settle within. This strategy is reasonable in a context such as human resettlement, where the settlers may try to combine the convenience of being close to others with the possible advantages associated with more space. We expect this resettlement scheme to yield distributions that are more homogeneous than those created by the $p\to 0$ limit of Model 1, and hence yield greater $D_q$.

We numerically obtain $D_q$ and find the dimension spectrum to be multifractal, having the same qualitative shape as the $p\to 0$ result for $D_q$ for Model 1. However,  due to the more uniform nature of the distributions created by this model, the specific values of $D_q$ are larger at each $q$ (e.g., $D_1=0.84\pm0.01$ [averaged over 20 distributions of $10^6$ points, where the error is the sample standard deviation] for this model versus our analytical result $D_1\simeq0.72$ for the $p\to 0$ limit of Model 1 [Eq.~\ref{d11}]).

\subsection{\label{m4}2-D Sparse Square (Model 6)}
\begin{figure}
\includegraphics[width=5in]{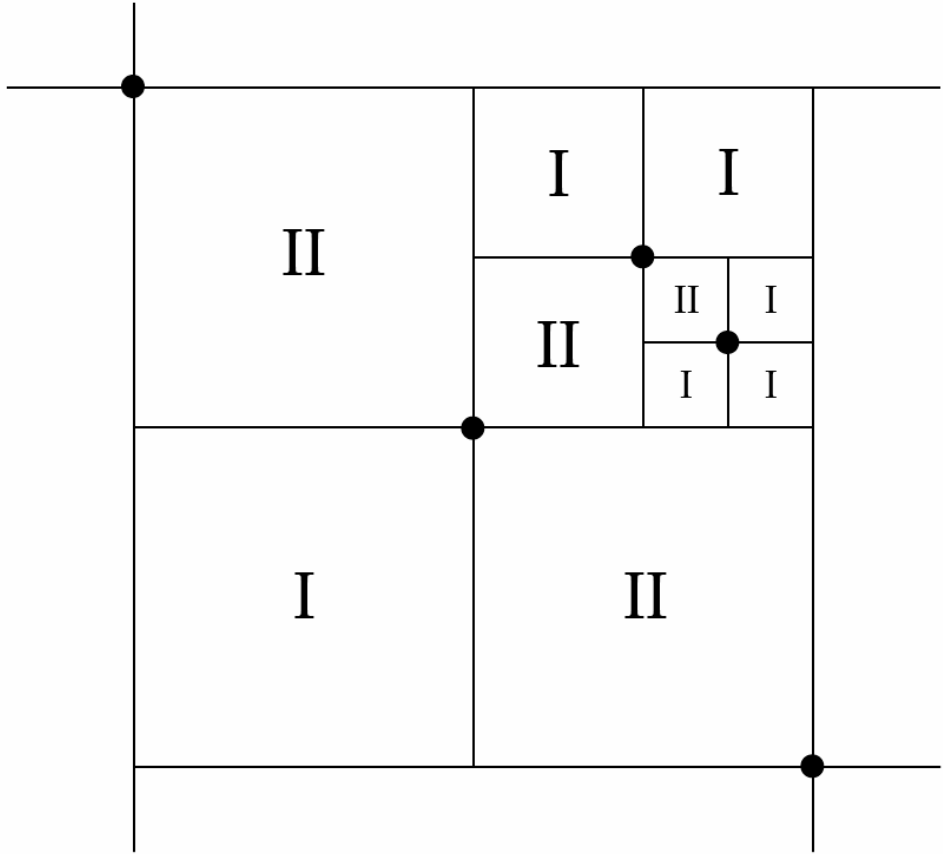}
\caption{\label{type12}Illustration of type I (labeled I) and type II (labeled II) squares, where black dots indicate populated vertices.}
\end{figure}

Model 6 is a two dimensional analogue of Model 5. The initial state and Step (a) of its construction are the same as in Model 4. Our aim is to extend Step (b) of Model 5, the choosing of the larger interval for settlement, to two dimensions, and, for that purpose, we make the following observation. In the square-based two dimensional point placement scheme of Model 4, there are two different types of squares which we refer to as type I and type II squares. A type I square only has one of its vertices occupied by a point, while a type II square has two opposite vertices occupied by points. (See Fig.~\ref{type12} for an illustration of type I and type II squares.) Thus a type I square can be regarded as more sparsely populated than a type II square of equal size. With this in mind, Step (b) for Model 6 is to place the new node in the middle of the largest square adjacent to the target point. In case of a tie, choose a type I square over a type II square, with any subsequent ties being resolved by an equal probability random choice from the remaining candidate squares.

\begin{figure}
\includegraphics[width=5in]{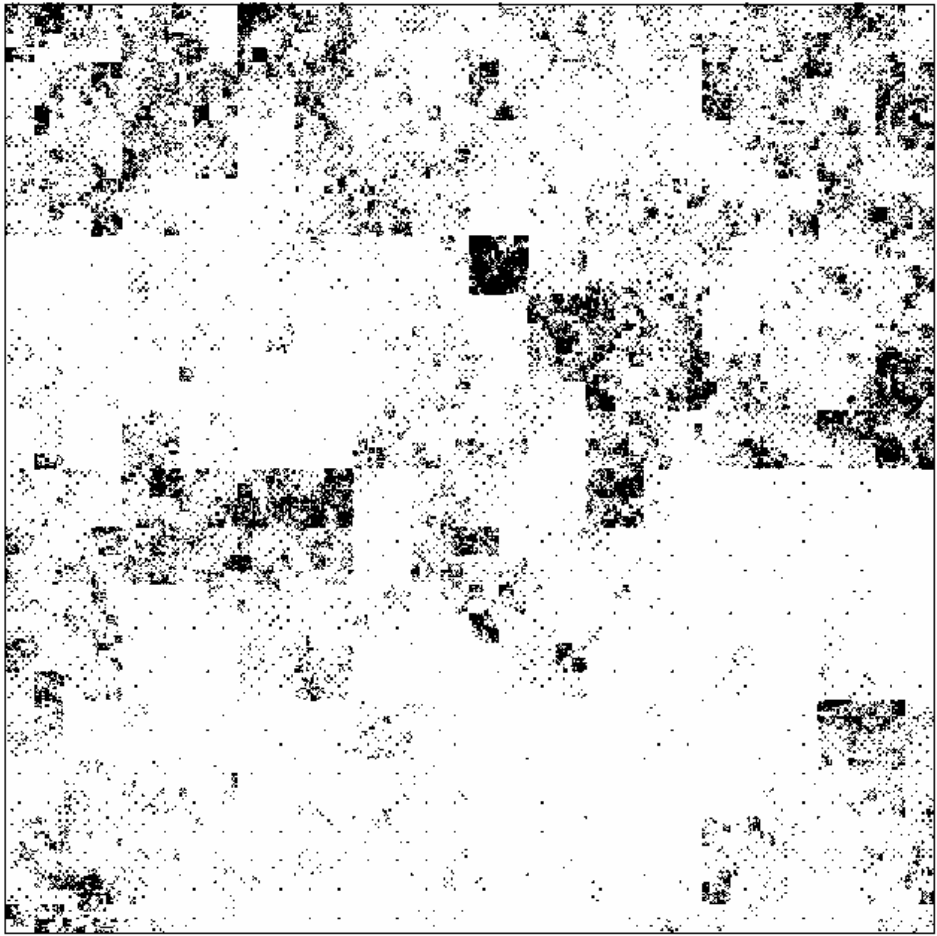}
\caption{\label{2dsp}A plot of a $10^5$ point distribution on the unit square generated by Model 6.}
\end{figure}

We plot a distribution generated by this model in Fig.~\ref{2dsp}. We find this distribution to be multifractal, with a $D_q$ that is similar in shape to, but everywhere larger than, the $D_q$ of Model 4 (except for $q=0$ where the two coincide with $D_0=2$). In particular, the information dimension is $D_1=1.30\pm0.02$ (averaged over 20 distributions of $10^6$ points), as compared to our analytical result of $D_1 \simeq 1.05$ for Model 4 (Eq.~\ref{d14}).

\subsection{2-D Unstructured (Model 7)}
Looking at Fig.~\ref{2dsq0p5} (Model 2), Fig.~\ref{2dsq} (Model 4) and Fig.~\ref{2dsp} (Model 6), we notice square shaped artifacts and diagonal point formations that arise due to the structured, square-based nature of those point placement schemes. Model 7 is a two dimensional point placement scheme that aims to avoid such artifacts, while maintaining the two ingredients of reproduction and local resettlement. We begin with a single point and assign to this point a {\it child distance} $d_c=1/4$.
At each subsequent discrete time step $t$, each preexisting point (parent) is selected to ``reproduce" with probability $p$, where the reproduction consists of a child point being placed at a random location on the circumference of a circle centered on the parent point and with radius equal to the parent point's $d_c$. A child distance equal to half the parent point's $d_c$ is assigned to the child point.

Model 7 results in multifractal point distributions for {\it all} values of $p$. This is in contrast to the trivial dimension spectrum $D_q=1$ for the $p=1$ limit of Model 1 and the trivial $D_q=2$ result for the $p=1$ limit of Models 2 and 3. In addition, while all the models presented thus far possess an integer valued box counting dimension, $D_0$, equal to the dimensionality of their respective ambient spaces, $D_0 \neq 2$ for all $p$ in Model 7.

\begin{figure}
\includegraphics[width=5in]{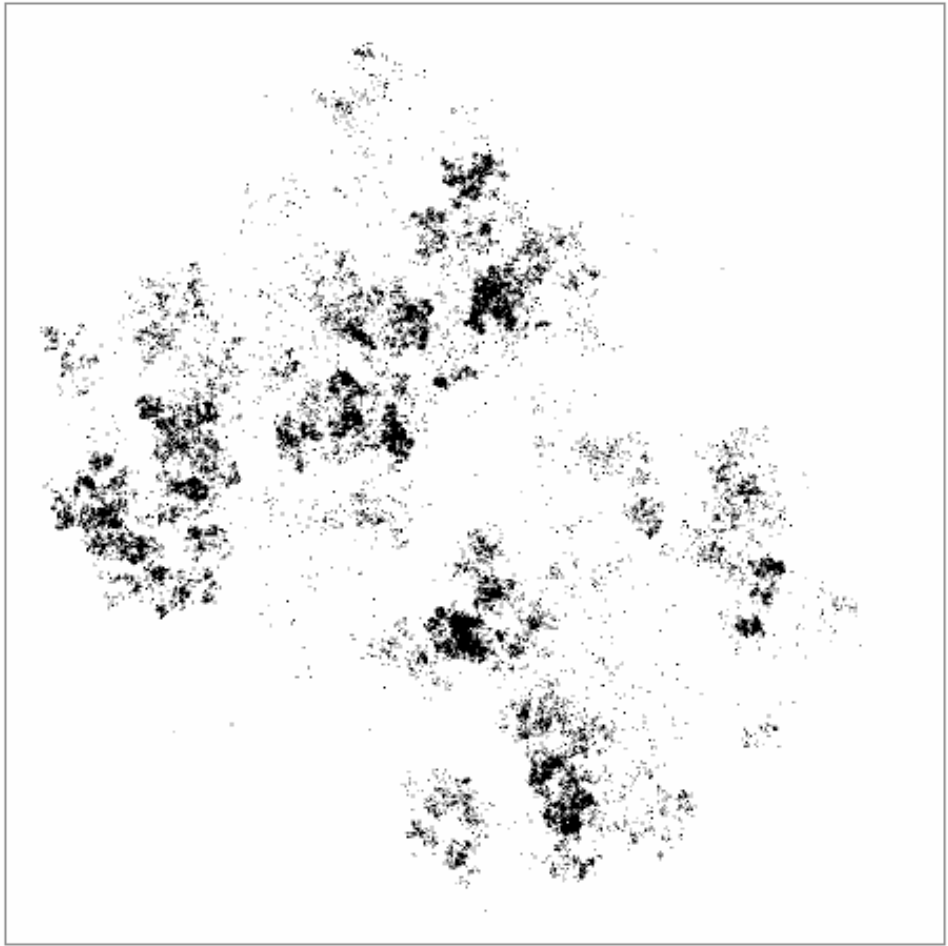}
\caption{\label{2dunst} A plot of a $10^5$ point distribution on the unit square generated by the $p\to 0$ construction of Model 7. }
\end{figure}

In a manner similar to the $p\to 0$ limits of Models 1 and 2, we can define the $p\to 0$ limit of Model 7 and create our point distribution one point at a time. The construction is as follows: At each time step a parent point is chosen from all preexisting points with equal probability. The parent point then produces a child according to the reproductive step described above in the general $p$ case. A distribution generated by the $p\to 0$ limit of Model 7 is shown in Fig.~\ref{2dunst}. As intended, there are no grid-type artifacts visible. The $p\to 0$ limit of Model 7 produces point distributions with $D_1 = 1.36 \pm 0.05$ (averaged over 20 distributions of $10^6$ points).

\subsection{Discussion}

Table~\ref{mt} summarizes our results for $D_1$ for our models and for the Earth at Night. Table~\ref{strat} summarizes gives qualitative descriptions of the reproduction and local resettlement mechanisms of our seven models.

 \begin{table}
 \caption{\label{mt} A summary of numerical and theoretical results for the point placement models and the Earth at Night image (Fig.~\ref{eatn}).}
 \begin{ruledtabular}
 \begin{tabular}{c l @{\extracolsep{\fill}} c @{\extracolsep{\fill}} c @{\extracolsep{\fill}} c}

 Model & & Dimensionality of Space & $D_1$ & $D_q$ (theory)\\
 \hline
 1 & & 1 & Eq.~(\ref{d11}) & Eq.~(\ref{dq1}) \\
 1 & ($p\to 0$) & 1 & $ 1 / (2 \ln 2)  \simeq 0.72$ & Eq.~(\ref{dq1p0}) \\
 2,3 & & 2 &  Eq.~(\ref{d123})  & Eq.~(\ref{dq23})\\
 2,3 & ($p\to 0$)& 2 & $ 3 / (4\ln 2) \simeq 1.08$ & Eq.~(\ref{dq2p0})\\
 4 & & 2 & $ 8 / (11 \ln 2) \simeq 1.05$ & Eq.~(\ref{dq4}) \\
 5 & & 1 & $0.84\pm0.01$ & - \\
 6 & & 2 & $1.30\pm0.02$ & - \\
 7 & ($p\to 0$) & 2 & $1.36\pm0.05$ & - \\
 \hline
  \multicolumn{2}{ c }{Earth at Night (North America)} & 2 & $1.65\pm0.02$ & -\\

 \end{tabular}
 \end{ruledtabular}
 \end{table}

\begin{table}
\caption{\label{strat}Summary of reproduction and local resettlement strategies employed in Models 1-7.}

\begin{tabular}{| c | l @{\hspace{1.2em}} |@{\hspace{1.5em}} l @{\hspace{1.5em}} |}\hline
\hspace{1em} Model \hspace{1em} & \hspace{1.2em} Reproduction & Resettlement \\
\hline\hline
1  & \hspace{1.2em} parent has one  & child settles in the middle of \\
  & \hspace{1.2em} \raisebox{1ex}[0pt]{child} &  \raisebox{1ex}[0pt]{chosen interval} \\
\hline
\raisebox{-1ex}[0pt]{2} & \hspace{1.2em} \raisebox{-1ex}[0pt]{parent reproduces} & \raisebox{-1ex}[0pt]{children share equally the}\\
  & \hspace{1.2em} \raisebox{.3ex}[0pt]{by four-fold} & \raisebox{.3ex}[0pt]{original square area occupied by}\\
  & \hspace{1.2em} \raisebox{1ex}[0pt]{cell division} & \raisebox{1ex}[0pt]{their parent} \\
\hline
 \raisebox{-1ex}[0pt]{3} & \hspace{1.2em} \raisebox{-1ex}[0pt]{parent has} & \raisebox{-1ex}[0pt]{parent remains in place;} \\
  &\hspace{1.2em}  \raisebox{.3ex}[0pt]{three children} & \raisebox{.3ex}[0pt]{children settle equal shares} \\
  & 			   & \raisebox{1ex}[0pt]{of parent's triangular area} \\
  \cline{2-3} 
  & \multicolumn{2}{| l |}{\raisebox{0.5ex}[0pt]{\hspace{1.5in} or}}\\
  \cline{2-3} 
  & \hspace{1.2em} \raisebox{-1ex}[0pt]{parent reproduces} & \raisebox{-1ex}[0pt]{children share equally the}\\
  & \hspace{1.2em} \raisebox{.3ex}[0pt]{by four-fold} & \raisebox{.3ex}[0pt]{original triangular area occupied by} \\
  & \hspace{1.2em} \raisebox{1ex}[0pt]{cell division} & \raisebox{1ex}[0pt]{their parent} \\
\hline
4 & \hspace{1.2em} parent has one & child settles randomly chosen\\
  & \hspace{1.2em} \raisebox{1ex}[0pt]{child} &  \raisebox{1ex}[0pt]{adjacent square} \\
\hline
5 & \hspace{1.2em} parent has one & child settles largest\\
  & \hspace{1.2em} \raisebox{1ex}[0pt]{child} &  \raisebox{1ex}[0pt]{adjacent interval} \\
\hline
6 & \hspace{1.2em} parent has one & child settles largest and\\
  & \hspace{1.2em} \raisebox{1ex}[0pt]{child} &  \raisebox{1ex}[0pt]{sparsest adjacent square} \\
\hline 
7 & \hspace{1.2em} parent has one & child settles\\
  & \hspace{1.2em} \raisebox{1ex}[0pt]{child} &  \raisebox{1ex}[0pt]{in a random direction} \\
\hline

\end{tabular}
\end{table}

Our result that the different strategies of reproduction and local resettlement used in these models (Table~\ref{strat}) all lead to multifractal spatial patterns of population density strongly suggests that multifractality may be a generic feature in real situations in which processes involving reproduction and local resettlement take place.

We emphasize that the spirit of our approach is very minimalist. In many real cases multiple interacting complex processes undoubtedly influence the determination of population patterns. For example, several important factors for the EaN image are geography (mountains, rivers, deserts, etc.), politics, societal and cultural factors, economics, etc. Our models show that considerations of underlying spatial heterogeneities are not required for explaining the existence of fractally heterogeneous distributions: even very simple dynamics incorporating reproduction and local resettlement are sufficient.

\section{\label{clip}The Effect of Clipped Data on Multifractality}

\begin{figure}
\includegraphics[width=5in]{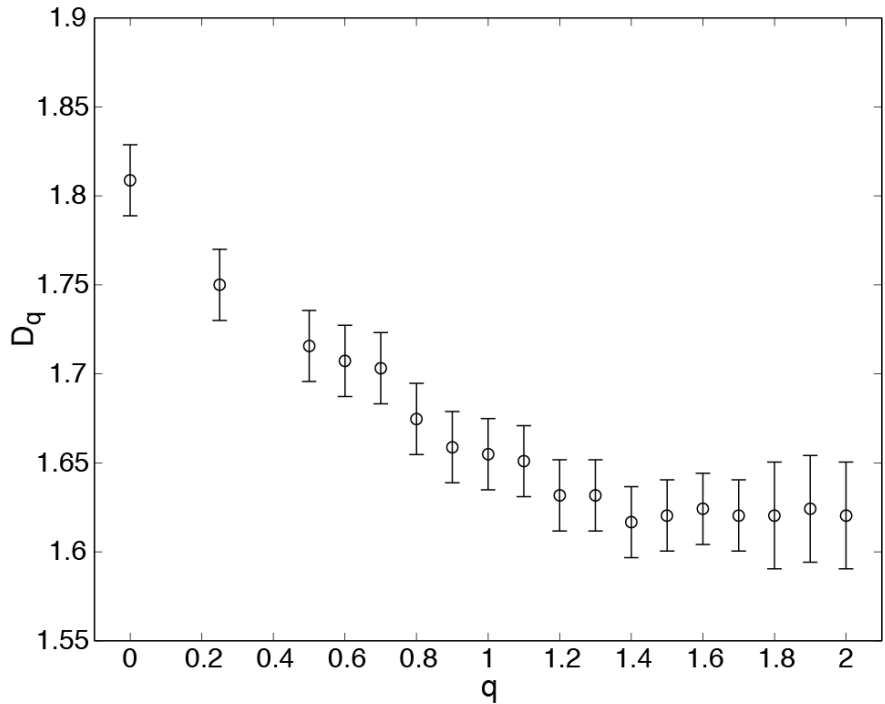}
\caption{\label{eatn_dq}Plot of $D_q$ for North America in the Earth at Night image. The error bars reflect the uncertainty involved in determining the slopes of the scaling regions in the plots of the quantity $(1-q)^{-1}\ln I(q,\epsilon)$ versus $\ln(1/\epsilon)$ [see Eq.~(\ref{dq})]. }
\end{figure}

We now revisit the North American part of the Earth at Night image (Fig.~\ref{eatn}) and analyze its fractal properties. Figure~\ref{eatn_dq} is a plot of $D_q$ calculated from the measured light distribution, where the error bars indicate the uncertainty involved in extracting the slopes of the various scaling regions. Although we observe a multifractal $D_q$ curve, studies on human population distributions~\cite{app1,app2} suggest that the a steeper $D_q$ should result. We now show how this discrepancy can be resolved.

\begin{figure}
\includegraphics[width=5in]{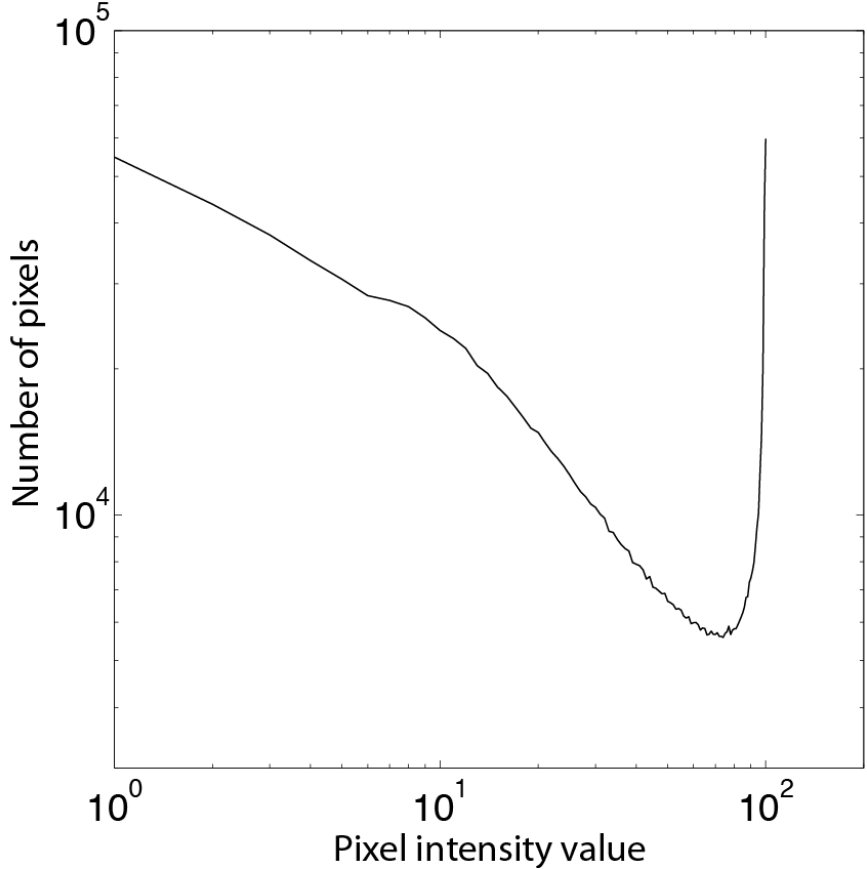}
\caption{\label{eatnhist}Histogram of individual pixel intensities for North America in the Earth at Night image.}
\end{figure}

Figure~\ref{eatnhist} is a log-log plot of the histogram of pixel intensities for North America in the EaN image. A spike is observed for a range of intensity values near the maximum value. This suggests that some regions with high light intensities caused some of the photoelectric cells of the satellites' sensors to saturate, thereby clipping the intensity values at a maximum allowable value. Additionally, it is known that, when a photoelectric cell is subjected to high intensities, it can trigger surrounding cells to register more light, resulting in the so called ``blooming" effect, which could account for the broadness of the observed high-intensity spike.  We note that the saturated pixels make up about $8\%$ of the total number of non-zero intensity pixels. We expect that blooming only changes intensities by a marginal amount, whereas clipping can make a drastic change. Thus the effect of blooming on $D_q$ should be negligible and we ignore it.

To illustrate the effects that this type of clipping has on multifractal distributions, we apply a scheme that mimics the clipping in the EaN image to a point distribution generated by one of our models. For this purpose we choose a $3\times10^6$ point distribution generated by the $p\to 0$ limit of Model 2. (Similar results are obtained using our other models.) First we roughly match the smallest scale of our distribution to that of the EaN image. Since the part of the EaN image we analyze is made up of $3000\times6000$ pixels, we cover our distribution with a fixed size grid of simulated pixels in which a simulated pixel corresponds to a grid square with edge length of $2^{-12}$, which is about $1/\sqrt{3000\times6000}$. Thus, while each pixel of the EaN image represents a light intensity value, each simulated pixel contains a number of individual points. The clipping procedure involves picking a maximum number of points allowed in a simulated pixel and changing the number of points in any simulated pixel containing more than this amount to the maximum amount. In order for the clipped point distribution to exhibit an appreciable range for the numbers of points in the individual simulated pixels we choose the maximum clipping value such that the total number of clipped simulated pixels make up only $4\%$ of the total nonempty simulated pixels (this is in contrast to the $8\%$ observed in the EaN image). For this particular distribution, any simulated pixel containing more than 40 points has its count reset to 40.

\begin{figure*}
\includegraphics[width=6.25in]{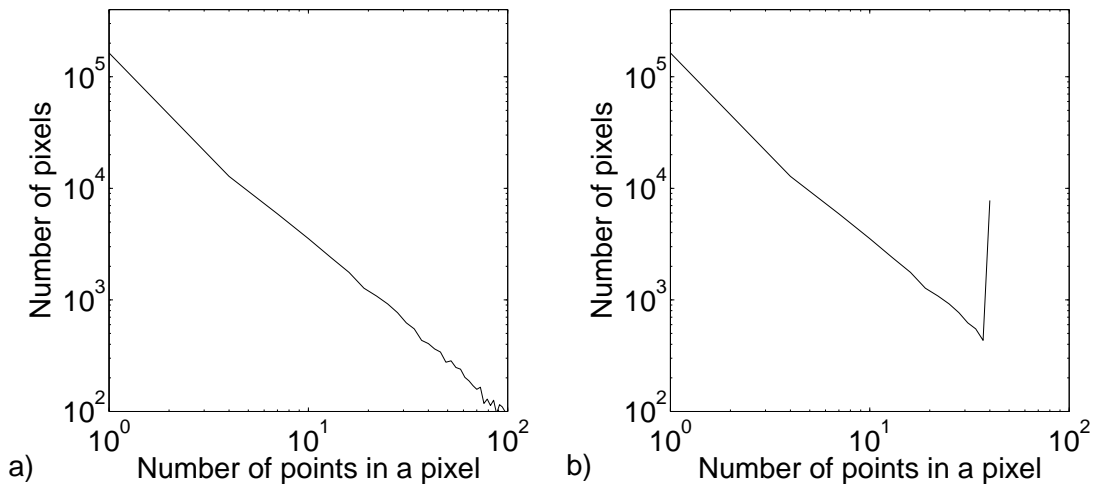}
\caption{\label{m2c_hist}Histograms of the number of points in individual simulated pixels (boxes in an $\epsilon=1/4096$ grid) used to cover a $3\times10^6$ point distribution generated by the $p\to 0$ limit of Model 2, a) for the original distribution and b) after the original distribution was clipped by allowing a maximum of 40 points in any simulated pixel.}
\end{figure*}

\begin{figure}
\includegraphics[width=6.25in]{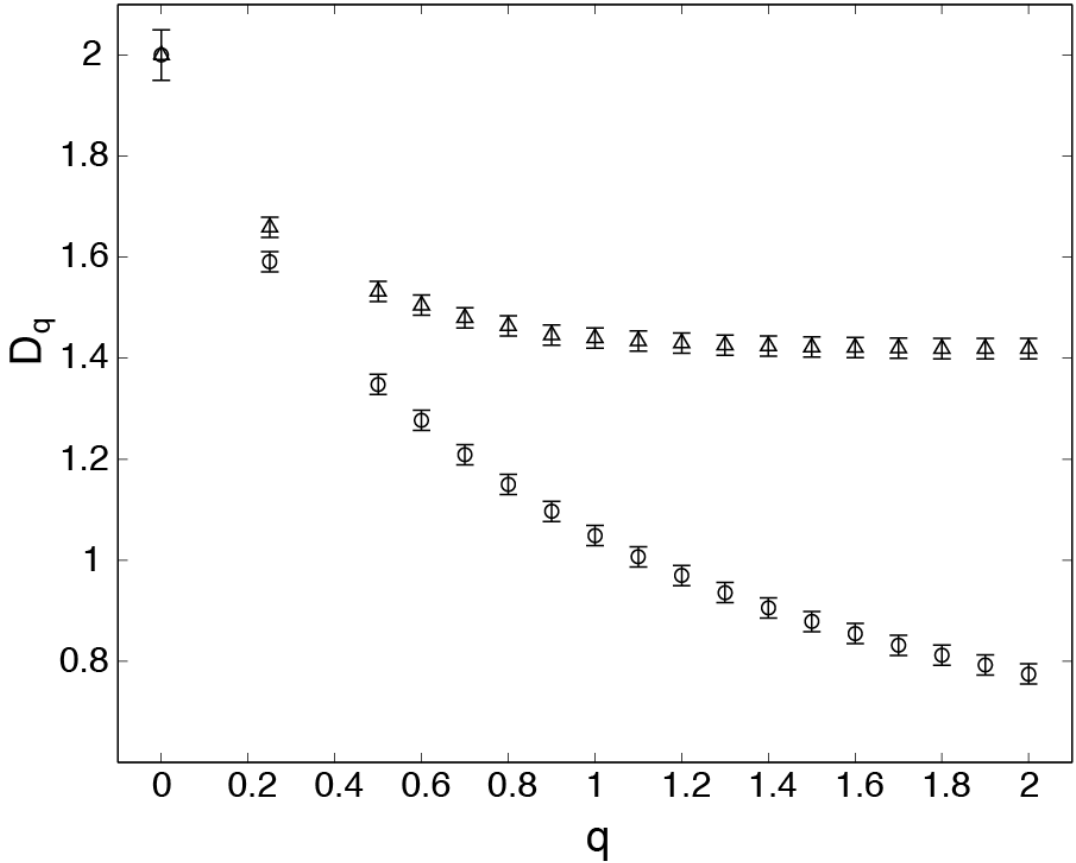}
\caption{\label{m2_clip_dq}Plot of $D_q$ for a $3\times10^6$ point distribution generated by the $p\to 0$ limit of Model 2 before the clipping procedure discussed in Section~\ref{clip} is applied (open circles), and after it is applied (open triangles). The error bars reflect the uncertainty involved in determining the slopes of the scaling regions in the plots of the quantity $(1-q)^{-1}\ln I(q,\epsilon)$ versus $\ln(1/\epsilon)$ [see Eq.~(\ref{dq})]. }
\end{figure}

Figures~\ref{m2c_hist}(a) and \ref{m2c_hist}(b) are histograms showing the distribution of the number of points contained in individual simulated pixels before and after the clipping is applied, respectively. We see that the clipped histogram is qualitatively similar to the histogram for the EaN image (Fig.~\ref{eatnhist}). Next, we calculate $D_q$ for the point distribution before and after clipping (open circles and open triangles in Fig.~\ref{m2_clip_dq}, respectively). We see that the clipping procedure transforms the unclipped $D_q$ to a flatter curve, making it more similar to the $D_q$ for North America in the EaN image (Fig.~\ref{eatn_dq}). 
Thus, we are able to conclude that the EaN image characteristics are consistent with a multifractal light intensity distribution with a steep $D_q$ sensed by an instrument that saturates at a maximum measurable intensity value.

\section{\label{inhom}Inhomogeneity}
Our models can be generalized to include geographical inhomogeneity. For example, one can imagine that the underlying space is supplemented by a fertility field, and that reproduction is more likely in regions of higher fertility. To investigate the effects of such inhomogeneities on the $D_q$ of a point distribution, we modify the $p\to 0$ construction of Model 2, by use of a fertility field $F(x,y)$ which we use to construct a space-dependent, parent node selection probability. If the location of point $i$ is $(x_i,y_i)$, then the selection probability for this point is taken to be $P_i = F(x_i,y_i) / \sum_j F(x_j,y_j)$ (this is in contrast to the original prescription of choosing the parent point from all preexisting points with equal probability).  We investigate two forms for the fertility field. First we use a smooth field
\begin{equation}
\label{Fs}
F_s(x,y)=1-0.3 \left\{ \cos (2\pi x) + \sin (2\pi y)\right\}
\end{equation}
on the unit square $(0 \leq x \leq 1, 0 \leq y \leq 1)$. We generate point distributions with this modified version of the $p\to 0$ limit of Model 2 and find that, while the regions of larger $F_s$ in the center of the square are much more dense with points than the regions of smaller $F_s$ near the corners, there is, nevertheless, no discernible difference in $D_q$ compared to the homogeneous case. That is, both the values of $D_q$, as well as the quality and extent of the scaling ranges, remain unchanged.

\begin{figure}
\includegraphics[width=4in]{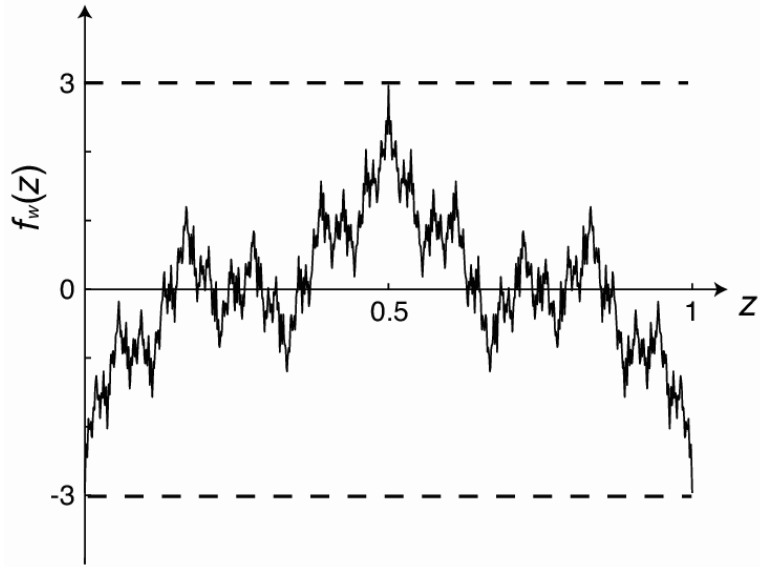}
\caption{\label{weier}A plot of the Weierstrass function $f_w(z)$ [Eq.~(\ref{fw})]. In our numerical implementation the infinite upper limit on the summation in Eq.~(\ref{fw}) is replaced by $j_{max}$, where $j_{max}$ is chosen such that the smallest value of $\epsilon$ used in determining $D_q$ is at least one order of magnitude larger than the smallest scale of the roughness, $3^{-j_{max}}$.}
\end{figure}

Next, to explore whether the above result can be attributed to the local smoothness of $F_s$,  we consider a rough fertility field,
\begin{equation}
\label{Fr}
F_r(x,y)=1+0.1 \left[ f_w(x) + f_w(y)\right],
\end{equation}
\begin{equation}
\label{fw}
f_w(z)\equiv-\sum_{j=0}^\infty \alpha^{-j} \cos (2\pi \beta^j z).
\end{equation}
with $\alpha =1.5$ and $\beta =3$. For $\alpha < \beta$, the function $f_w(z)$ is a `Weierstrass function' (see Fig.~\ref{weier}); it is rough in the sense that, although it is continuous, it is nondifferentiable, and the graph of $f_w(z)$ versus $z$ is a fractal curve (the fractal dimension of the curve is $2-(\ln \alpha)/(\ln \beta) \simeq 1.63$~\cite{eo}). With $\alpha =1.5$, $F_r$ has the same range of variation as $F_s$ (i.e., 0.4 to 1.6). Applying this fertility field to the $p\to 0$ limit of Model 2 results in the same values and scaling ranges for $D_q$ as in the homogeneous case and the smooth fertility field case [Eq.~(\ref{Fs})]. This suggests that, while inhomogeneities of the underlying space may dictate certain aspects of the distribution of a growing population, for example, where the points are more likely to settle, its fractality can be due mainly to reproduction and local resettlement processes.

\section{\label{theory}Theory}
\subsection{\label{theorym1}Model 1}
Here we derive an analytical expression for the fractal dimension spectrum $D_q$ of population distributions generated by the one dimensional point placement scheme of Model~1.

We define $N(k,t)$ to be the average (over different realizations of Model 1) number of intervals of length $2^{-k}$ at time $t$. The initial condition for this model (at $t=1$) is a single point on the unit length circumference of a circle, implying $N(k,1) = \delta_{k 0}$, where $\delta_{k j}$ is the Kronecker delta. At each time $t$, each preexisting interval is bisected with probability $p$. Thus, with probability $p$ an interval of length $2^{-k}$ is replaced by two intervals of lengths $2^{-(k+1)}$. The discrete time evolution equation for $N$ is then
\begin{equation}
\label{Nkt}
N(k,t+1) = 2 p N(k-1,t) + (1-p) N(k,t),
\end{equation}
where $p N(k-1,t)$ is the average (over different realizations of Model 1) number of intervals of length $2^{-(k-1)}$ bisected at time $t+1$, and $(1-p) N(k,t)$ is the average number of $2^{-k}$ length intervals that remain unchanged at $t+1$. We now define $N(t)= \sum_k N(k,t)$ to be the average number of intervals (of all sizes) at time $t$. Taking the sum over $k$ of Eq.~(\ref{Nkt}) we obtain the simple recursion relation
\begin{equation}
\label{Nt}
N(t+1)=(1+p) N(t),
\end{equation}
which, since $N(1)=1$, gives the result $N(t)=(1+p)^{t-1}$ for the evolution of the average total number of intervals.

In Appendix~\ref{pf}, we introduce the partition function formalism for calculating $D_q$. We define the quantity $\tau = (q-1) D_q$ and derive the following expression for $q(\tau)$ for Model 1 [Eq.~(\ref{A_qtau})]:

\begin{equation}
\label{qtau}
q(\tau) = 1+\lim_{t\to \infty} { \ln \langle 2^{k \tau} \rangle_t \over \ln N(t)},\; \; \; \langle 2^{k \tau} \rangle_t = {\sum_{k} N(k,t) 2^{k \tau} \over N(t)}.
\end{equation}

Multiplying both sides of Eq.~(\ref{Nkt}) by $2^{k \tau}$ and then summing over $k$, we obtain the recursion relation,
\begin{equation}
\langle 2^{k \tau} \rangle_{t+1} = {1 \over 1+p}  \left( 2^\tau 2 p + 1 - p\right) \langle 2^{k \tau} \rangle_t,
\end{equation}
which, since $\langle 2^{k \tau} \rangle_{t=1} = 1$, gives,
\begin{equation}
\label{2ktau}
 \langle 2^{k \tau} \rangle_t = \left({1 \over 1+p}  \left( 2^\tau 2 p + 1 - p\right) \right)^{t-1}.
\end{equation}
Substituting Eq.~(\ref{2ktau}) into Eq.~(\ref{qtau}), we obtain
\begin{equation}
\label{q}
q={ \ln \left(2^\tau 2 p + 1 - p\right) \over \ln ( 1 + p) }.
\end{equation}
Then, since $D_q = \tau(q)/(q-1)$, we invert Eq.~(\ref{q}) for $\tau(q)$ and obtain the following result for $D_q$:
\begin{equation}
\label{dq1}
D_q = { \log_2 \left( {p - 1 + (1 + p)^q \over 2 p}\right) \over q-1}.
\end{equation}
In particular, using L'Hospital's Rule,
\begin{equation}
\label{d11}
D_1 = { (1+p) \log_2 (1+p) \over 2 p }.
\end{equation}
For $p=1$ we get the expected result $D_q = 1$, while for the $p\to 0$ limit, we obtain
\begin{equation}
\label{dq1p0}
D_q = { \log_2 (1 + q) - 1 \over q-1}
\end{equation}
and
\begin{equation}
\label{d11p0}
D_1 = { 1 \over 2 \ln 2 } \simeq 0.72.
\end{equation}

\subsection{\label{theorym23}Models 2 and 3}
Here we derive the analytical $D_q$ for Models 2 and 3. We adopt the area partitioning viewpoint presented in Sec.~\ref{m6}, where we focus on the squares containing the points rather than the points themselves.
We begin by defining $N(k,t)$ to be the average number of squares with edge length $2^{-k}$ at time $t$. 
The initial state (at $t=1$) is a single square (with a point at its center) with unit edge length, giving us $N(k,1)=\delta_{k 0}$. At each time $t$, each preexisting square is divided into four equal sized squares with half the edge length of the original square with probability $p$ (Fig.~\ref{sqfig}). Thus, with probability $p$ a square of edge length $2^{-k}$ is replaced by four squares of edge lengths $2^{-(k+1)}$. The discrete time evolution equation for $N(k,t)$ is then
\begin{equation}
\label{Nkt2}
N(k,t+1) = 4 p N(k-1,t) + (1-p) N(k,t).
\end{equation}
We define $N(t) = \sum_k N(k,t)$ to be the average number of squares (of all sizes) at time $t$, and take the sum over $k$ of Eq.~(\ref{Nkt2}) to obtain
\begin{equation}
\label{Nt2}
N(t+1)=(1+ 3 p) N(t),
\end{equation}
which, since $N(1)=1$, gives the result $N(t)=(1+ 3 p)^{t-1}$ for the evolution of the average total number of intervals.
Interchanging the squares in the above derivation with equilateral triangles, one can see that the same evolution equation holds for the triangles in Model 3.

In Appendix~\ref{pf} we show that Eq.~(\ref{qtau}), which allowed us to calculate $q(\tau)$ in terms of $\langle 2^{k \tau} \rangle_t $ for Model 1, holds for Models 2 and 3 as well. Following the procedure in Sec.~\ref{theorym1}, we multiply both sides of Eq.~(\ref{Nkt2}) by $2^{k \tau}$ and sum over $k$, solving the resulting recursion relation to obtain
\begin{equation}
\label{2ktaum67}
 \langle 2^{k \tau} \rangle_t = \left({1 \over 1+ 3 p}  \left( 2^\tau 4 p + 1 - p\right) \right)^{t-1}.
\end{equation}
Substituting Eq.~(\ref{2ktaum67}) into Eq.~(\ref{qtau}), we obtain
\begin{equation}
\label{q2}
q={ \ln \left(2^\tau 4 p + 1 - p\right) \over \ln ( 1 + 3 p) }.
\end{equation}

Thus, for Models 2 and 3 we obtain the multifractal dimension spectrum
\begin{equation}
\label{dq23}
D_q = { \log_2 \left( {p - 1 + (1 + 3 p)^q \over 4 p}\right) \over q-1}.
\end{equation}
and the information dimension
\begin{equation}
\label{d123}
D_1 = { (1+ 3 p) \log_2 (1+ 3 p) \over 4 p }.
\end{equation}

For $p=1$ we get $D_q = 2$, while for the $p\to 0$ limit, we obtain
\begin{equation}
\label{dq2p0}
D_q={\log_2 ( 1 + 3 q ) - 2 \over q-1}
\end{equation}
and
\begin{equation}
\label{d12p0}
D_1={3 \over 4\log 2}\simeq 1.08.
\end{equation}

\subsection{\label{theorym4}Model 4}
The calculation of $D_q$ for Model 4 is based on the classification of squares into types I and II, discussed in Sec.~\ref{m4}. We begin by defining $N_1(k,t)$ and $N_2(k,t)$ to be the average (over different realizations) number of type I and type II squares, respectively, with edge length $2^{-k}$ at time $t$. The initial state ($t=1$) is a single point at $(0.5,0.5)$. The point divides the unit square into four equal squares of edge length $1/2$, with each square having one point on one of its vertices.  Thus $N_1(k,1)=4 \delta_{k 1}$, while $N_2(k,1)=0$ for all $k$. At each time $t$, the new point is equally likely to appear in each type I square, but is twice as likely to appear in each type II square, because there are two parent points that can produce offspring in each type II square. In other words,  a type I square with edge length $2^{-k}$ is chosen with probability $N_1(k,t)/\sum_j \left( N_1(j,t) + 2 N_2(j,t) \right)$, while a type II square of the same size is chosen with probability $2 N_2(k,t)/\sum_j \left( N_1(j,t) + 2 N_2(j,t) \right)$. When a type I square is chosen, the new point is placed in its center, destroying the original square and creating three type I squares and one type II square, all with half the edge length of the original square. On the other hand, when the new point is in a type II square, it is replaced by two type I squares and two type II squares, all having half the edge length of the original square. Notice that either way, the quantity $\sum_j  \left( N_1(j,t) + 2 N_2(j,t) \right)$ increases by 4 between time $t$ and $t+1$, so that $\sum_j  \left( N_1(j,t) + 2 N_2(j,t) \right)= 4 t$.

The discrete time evolution equations for both types of squares are then
\begin{equation}
\label{N1}
N_1(k,t+1) = N_1(k,t)\left(1-{1 \over 4 t} \right) + N_1(k-1,t) {3 \over 4 t} + 2 N_2(k-1,t) {2 \over 4 t},
\end{equation}
\begin{equation}
\label{N2}
N_2(k,t+1) = N_2(k,t)\left(1-{2 \over 4 t} \right) + N_1(k-1,t) {1 \over 4 t} + 2 N_2(k-1,t) {2 \over 4 t}.
\end{equation}
In Appendix~\ref{pf} we show that for Model 4 [Eq.~(\ref{A_qtau2})],
\begin{equation}
\label{qtau2}
q(\tau) = \lim_{t\to \infty} { \ln \left(T_1(t) + 2^q T_2(t)\right) \over \ln t},
\end{equation}
where $T_i(t)=\sum_k 2^{ k \tau} N_i(k,t)$.
Multiplying Eqs.~(\ref{N1}) and (\ref{N2}) by $2^{k \tau}$ and summing over $k$, we obtain
\begin{eqnarray}
\label{T1}
T_1(t+1) & = & T_1(t)\left( 1+ {3\cdot2^\tau -1 \over 4 t} \right) + T_2(t){4\cdot2^\tau \over 4 t} ~\\
\label{T2}
T_2(t+1) & = & T_2(t)\left( 1+ {4\cdot2^\tau -2 \over 4 t} \right) + T_1(t){2^\tau \over 4 t}.
\end{eqnarray}

At this point, we make the continuous time approximation $T_i(t+1)-T_i(t)\approx {\mathrm d }T_i(t)/{\mathrm d} t$, valid for large $t$, which allows us to represent Eqs.~(\ref{T1}) and (\ref{T2}) as the matrix differential equation,
\begin{equation}
\label{Ts}
\frac{{\mathrm d} \mathbf T(s)}{{\mathrm d} s} = \mathbf M \cdot \mathbf T(s),
\end{equation}
where
\begin{equation}
\mathbf T(s) = \left[\begin{array}{c}
T_1(s)\\
T_2(s)
\end{array}
 \right],
 \end{equation}
 \begin{equation}
 \mathbf M = \left[ \begin{array}{cc}
3\cdot 2^\tau - 1 & 4\cdot 2^\tau \\
2^\tau & 4\cdot 2^\tau - 2 \end{array}
\right]
\end{equation}
and $s=(1/4)\ln t$.

Equation~(\ref{Ts}) is solved in terms of the eigenvalues $\lambda_a,\lambda_b$ and the eigenvectors $\mathbf T_{a},\mathbf T_{b}$ of the matrix $\mathbf M$, giving
\begin{equation}
\mathbf T_i \sim e^{\lambda_i s} = t^{\lambda_i \over 4},
\end{equation}
for $i=a,b$, with $\lambda_a,\lambda_b$ distinct and real and $\lambda_a>\lambda_b$.
We can express the quantity $(T_1(t)+2^qT_2(t))$ in Eq.~(\ref{qtau2}) in terms of a linear combination of the components of these eigenvectors, whose long time behavior is dominated by the behavior of $\mathbf T_a$, the eigenvector associated with the larger eigenvalue $\lambda_a$. Thus (\ref{qtau2}) gives us the simple result
\begin{equation}
\label{qlambda}
q=\lambda_a/4.
\end{equation}

The eigenvalues of $\mathbf M$ are the roots of its characteristic polynomial,
\begin{equation}
\lambda^2 + \lambda (3-7\cdot2^\tau) + 2 -10\cdot 2^\tau + 8 \cdot 2^{2 \tau} = 0.
\end{equation}
This equation is also quadratic in $2^\tau$, allowing us to solve for $\tau(\lambda)$
\begin{equation}
\label{taulambda}
\tau(\lambda)=\log_2 \left[ {1 \over 16} \left(10 + 7 \lambda \pm \sqrt{17 \lambda^2 + 44 \lambda + 36 } \right) \right].
\end{equation}
We substitute the relationship from Eq.~(\ref{qlambda}) into Eq.~(\ref{taulambda}) and, since $\tau(q)=(q-1)D_q$, we resolve the ambiguity of the $\pm$ sign by requiring that $\tau(q=1)=0$. This results in
\begin{equation}
\label{dq4}
D_q={1 \over q-1} \log_2 \left[ {1 \over 8} \left(5 + 14 q - \sqrt{68 q^2 + 44 q + 9 } \right) \right]
\end{equation}
for the dimension spectrum of Model 4.
In particular, applying L'Hospital's Rule to (\ref{dq4}), the information dimension is
\begin{equation}
\label{d14}
D_1={8 \over 11 \ln 2}\simeq 1.05.
\end{equation}

\section{\label{conclusion}Conclusion}

The main conclusion of this paper is that reproduction and local resettlement processes may lead to multifractal spatial distributions for growing populations. We introduced a number of point placement models in one and two dimensions and showed that the models resulted in multifractal distributions. Furthermore, we have demonstrated qualitative similarity between the example of the Earth at Night image and clipped versions of distributions generated by our models. We thus suggest that the mechanism by which our models create multifractal distributions may be operative in the growth of real systems.

We thank Dan Lathrop and Rowena Ball for useful conversations. This work was supported by ONR (Physics) and by NSF (Contract Nos. PHYS 0098632 and DMS 0104087).

\appendix
\section{\label{pf}Calculation of Model Dimension Spectra}
\subsection{\label{pf1}The Partition Function Formalism}
The partition function formalism~\cite{grass2,halsey} is an alternative to the fixed sized grid method of Eq.~(\ref{dq}) for calculating a dimension spectrum of a measure. We demonstrate here how it allows us to relate the various interval and square size distribution functions found in Models 1, 2, 3 and 4 to the dimensions of the point distributions generated by them.

We cover the measure of interest with a disjoint covering $\{S_i\}$, $i=1,2...N$, where each element $S_i$ of the covering set has a diameter $\epsilon_i$ less than or equal to $\delta$. (The diameter $\epsilon_i$ is the largest possible distance between two points in $S_i$.) The partition function is defined as
\begin{equation}
\label{gammaq}
\Gamma_q(\tau,\{S_i\}, \delta)= \sum_{i=1}^N \mu_i^q/\epsilon_i^{\tau},
\end{equation}
where $\mu_i$ is the measure of $S_i$.
For a given $\delta$ the covering $\{S_i\}$ is now chosen such that Eq.~(\ref{gammaq}) is maximized (for $q> 1$) or minimized (for $q< 1$), which defines
\begin{equation}
\label{optcov}
\Gamma_q(\tau, \delta)=
\left\{
\begin{array}{ccc}
\sup_{S_i} \Gamma_q(\tau, \{S_i\}, \delta),&{\rm for} \  q>1& \\
\inf_{S_i} \Gamma_q(\tau, \{S_i\}, \delta),&{\rm for} \  q<1&.
\end{array}
\right.
\end{equation}
Then, letting $\delta\to0$, we define
\begin{equation}
\label{gtauapp}
\Gamma_q(\tau)=\lim_{\delta\to0} \Gamma_q(\tau, \delta).
\end{equation}
The quantity $\Gamma_q(\tau)$ experiences a jump from 0 to $+\infty$, as $\tau$ is increased, at a critical value that we denote $\tau(q)$. The dimension $\tilde D_q$ of the measure is then defined as
\begin{equation}
\label{dqtilde}
\tilde D_q=\tau(q)/(q-1).
\end{equation}

In practice, it is difficult to determine whether a particular covering $\{ S_i \}$ is optimal in the sense of Eq.~(\ref{optcov}). However, one can often compute the correct dimension by considering specific coverings, in the following sense. Consider a sequence of disjoint coverings $\{S_i^{(m)}\}$ where covering $m$ of the sequence has maximum diameter $\delta^{(m)}$ ($\epsilon_i^{(m)}<\delta^{(m)}$ for every component $i$ of covering $m$), and $\delta^{(m)}$ converges to zero as $m$ approaches infinity.
If the limit 
\begin{equation}
\label{gq}
\hat \Gamma_q(\tau) = \lim_{m\to\infty} \Gamma_q(\tau,\{S_i^{(m)}\}, \delta^{(m)})
\end{equation}
exists, then like $\Gamma_q(\tau)$ it experiences a jump from 0 to $+\infty$, as $\tau$ increases, at a critical value $\tau = \hat \tau (q)$. In terms of the sequence $\{ S_i^{(m)} \}$ of coverings (not necessarily optimal), we can then compute a dimension spectrum
\begin{equation}
\hat D_q=\hat \tau(q)/(q-1),
\end{equation}

Since the sequence of coverings used to compute $\hat D_q$ may be suboptimal, we have $\hat \tau(q) \geq \tau(q)$ for $q>1$, and $\hat \tau(q) \leq \tau(q)$ for $q<1$. Thus $\hat D_q \geq \tilde D_q$ in each case. Assuming the limit in Eq.~(\ref{dq}) that defines $D_q$ to exist, then $D_q$ can be computed by the partition function formalism above using equal size coverings ($\epsilon_i^{(m)} = \delta^{(m)}$ for every $i$ and $m$). Thus $D_q \geq \tilde D_q$ as well. While there do exist examples for which $\tilde D_q \not = D_q$ for $q<1$ \footnote{As a simple example, we consider the set, $1,1/2,1/3,\cdots,n^{-1},\cdots$, and $q=0$. For this set it can be shown that Eqs.~(\ref{gammaq})-(\ref{dqtilde}), yield $\tilde D_0 = 0$, which is intuitively reasonable for a set that is a countable collection of points. In contrast, Eqs.~(\ref{dq}) and (\ref{Iq}) yield $D_0 = 1/2$. To see this we first note that, for $q=0$, (\ref{dq}) and (\ref{Iq}) give the well-known result $D_0 = \lim_{\epsilon\to0}\{[\ln N(\epsilon)]\left/[\ln (1/\epsilon)]\right.\}$, where $N(\epsilon)$ is the number of $\epsilon$-intervals needed to cover the set. Next we observe that the distance between $1/n$ and $1/(n+1)$ is approximately $1/n^2$ for large $n$. Thus setting $\epsilon = 1/n^2$, we need $n$ intervals to cover the first $n$ elements of the set, one interval for each such element. To cover the remaining elements we must cover the interval $(1/(n+1),0)$. This requires $(1/\epsilon)[1/(n+1)]=n^2/(n+1)\simeq n$ intervals. Thus $N(\epsilon)\simeq 2 n = 2/\epsilon^{1/2}$, yielding $D_0=1/2$.}
the two dimensions are typically found to coincide in analytical examples with physical bases. Furthermore, for $q>1$ the two dimensions always coincide:

{\bf Theorem}: For every probability measure $\mu$ for which the limit in Eq.~(\ref{dq}) exists, $\tilde D_q = D_q$ for all $q>1$. (If one defines $D_q$ as a liminf, then this equality holds for all $\mu$.)\\
We prove this theorem in Appendix~\ref{ec}.

The theorem demonstrates that, for $q>1$ at least, the dimension spectrum computed according to the partition function formalism from a particular sequence of coverings is not that sensitive to the type of covering; equal size coverings yield the same spectrum as a sequence of coverings that is optimal in the sense of Eq.~(\ref{optcov}). While it may be possible to compute a different value of $\hat D_q$ from a sufficiently suboptimal sequence of coverings, we conjecture that for the coverings we consider below,
\begin{equation}
\label{dcalc}
\hat D_q = \tilde D_q = D_q.
\end{equation}
Furthermore, we conjecture that Eq.~(\ref{dcalc}) holds not only for $q>1$, but for all $q\geq 0$.

For our purposes, the application of Eq.~(\ref{dcalc}) for $q\leq1$ is supported by the following reasoning. In all of the cases below, the formula we obtain for $\hat D_q$ is an analytic function of $q$ (for our region of interest $0\leq q \leq 2$). If $\hat D_q=D_q$ for $q>1$, it follows that if $D_q$ is continuous at $q=1$, then $\hat D_1=D_1$. Furthermore, if $D_q$ is analytic, then $\hat D_q=D_q$ for $q<1$ as well.

\subsection{\label{pf2}Application to the Models}

The practical implication of Eq.~(\ref{dcalc}) for us is significant. It means that we can choose any sequence of coverings whose maximum diameter converges to zero, and this choice can be made in a manner that {\it facilitates analytic computations}. The result will be the same as that for an equal cube size covering [Eq.~(\ref{dq})] or an optimal covering [Eq.~(\ref{optcov})].

For Model 1, we choose the intervals between the points at time $t$ as a covering for the distribution. We regard each interval as covering the equivalent of one point since it contains two half points at each of its ends. Hence, on average, $\mu_i=(\sum_k N(k,t))^{-1}=1/N(t)$ for all $i$, and Eq.~(\ref{gq}) becomes \footnote{There is a subtlety here. As $t$ increases we are changing {\it both} the covering {\it and} the distribution of points. We really should be looking at the limiting distribution and not the time $t$ distribution. Since each interval is equally likely to subdivide, the {\it expected} measure in each of the time $t$ intervals is the same, and our computations in this section reflect this approximation.}
\begin{equation}
\label{gq1}
\hat \Gamma_q(\tau)=\lim_{t\to\infty} (1/N(t))^q \sum_{i} \epsilon_i^{-\tau},
\end{equation}
where the $\delta\to0$ limit is replaced by $t\to\infty$, since, as $t$ increases the size of the largest interval decreases to zero with probability one. Since $\ln \hat \Gamma_q(\tau)= -\infty$ for $\tau < \hat\tau(q)$ and $\ln \hat \Gamma_q(\tau)= \infty$ for $\tau > \hat\tau(q)$, we equate $\ln \hat \Gamma_q(\tau)$ to zero and obtain $q$ as a function of the transition value $\hat \tau$,
\begin{equation}
q(\hat \tau)=\lim_{t\to\infty} {\ln \sum_{i} \epsilon_i^{-\hat \tau} \over \ln N(t)}=\lim_{t\to\infty} {\ln \sum_{k} N(k,t) 2^{k \hat \tau} \over \ln N(t)}.
\end{equation}
We then have [see Eq.~(\ref{qtau})]
\begin{equation}
\label{A_qtau}
q(\hat \tau) = 1+\lim_{t\to \infty} { \ln \langle 2^{k \hat \tau} \rangle_t \over \ln N(t)},
\end{equation}
where
\begin{equation}
\langle 2^{k \tau} \rangle_t = {\sum_{k} N(k,t) 2^{k \tau} \over N(t)}.
\end{equation}

Each square in Model 2 and triangle in Model 3 contains, on average, $(\sum_k N(k,t))^{-1}=1/N(t)$ of the total measure. Thus, like we derived for Model 1, we obtain
\begin{equation}
\label{A_qtau_23}
q(\hat \tau) = 1+\lim_{t\to \infty} { \ln \langle 2^{k \hat \tau} \rangle_t \over \ln N(t)}.
\end{equation}

For Model 4 we use the two types of squares (type I and II) to cover the distribution. A type I square has one point on one of its vertices and hence covers the equivalent of a quarter of a point, while a type II square, with two points on its vertices, covers the equivalent of half a point. Thus, since at time $t$ we have $t$ total points, the measure contained in a type I square is $1/4 t$, while that contained in a type II square is $1/2 t$. Thus Eq.~(\ref{gq}) becomes
\begin{equation}
\label{gq2}
\hat \Gamma_q(\tau)=\lim_{t\to\infty} {1 \over (4 t)^q} \left( \sum_{i = 1}^{N_1} \epsilon_i^{-\tau} + 2^q \sum_{i = 1}^{N_2} \epsilon_i^{-\tau} \right),
\end{equation}
where the first summation is over all type I squares and the second is over all type II squares.
Taking the logarithm of this expression and equating it to zero, we obtain
\begin{equation}
\label{A_qtau2}
q(\hat \tau) = \lim_{t\to \infty} { \ln \left(T_1(t) + 2^q T_2(t)\right) \over \ln t},
\end{equation}
where $T_i(t)=\sum_k 2^{ k \hat \tau} N_i(k,t)$.

\section{\label{ec}Proof of Theorem from Appendix~\ref{pf}}

As we noted in Appendix~\ref{pf}, $\tilde D_q \leq D_q$ for all $q\geq 0$, so we now show that also $\tilde D_q \geq D_q$ for $q>1$. Our proof is based on the theorem in \cite{huntkaloshin} that for $q>1$, the quantity
\begin{equation}
I_q(s)=\int \left( \int {d\mu(y) \over \left| x-y\right|^s}\right)^{q-1} d\mu(x)
\end{equation}
is finite for $s<D_q$ and infinite for $s>D_q$.
To see that $\tilde D_q \geq D_q$, we show that $\tilde D_q \geq s$ whenever $I_q(s)$ is finite. Let $\{S_i\}$ be a disjoint covering of $\mu$, as in Appendix \ref{pf}, and recall that $\mu_i$ is the measure of $S_i$ and $\epsilon_i$ is its diameter. Then considering only the contribution to $I_q(s)$ from points $x$ and $y$ that are in the same $S_i$, we have
\begin{eqnarray}
I_q(s) & \geq & \sum_{i=1}^N \int_{S_i} \left( \int_{S_i} {d\mu(y) \over \left| x-y\right|^s}\right)^{q-1} d\mu(x) \nonumber \\
& \geq & \sum_{i=1}^N \int_{S_i} \left( \int_{S_i} {d\mu(y) \over \epsilon_i^s}\right)^{q-1} d\mu(x) \nonumber \\
& = & \sum_{i=1}^N \mu_i \left( \mu_i \over \epsilon_i ^s \right)^{q-1} \nonumber \\
& = & \sum_{i=1}^N \mu_i ^q \left/ {\epsilon_i}^{(q-1) s} \right. \nonumber \\
& = & \Gamma_q \left( (q-1) s, \{ S_i \}, \delta \right) \nonumber 
\end{eqnarray}
by (\ref{gammaq}). This implies that $\Gamma_q \left( (q-1) s, \delta \right) \leq I_q(s)$ by (\ref{optcov}), since $q>1$, and hence $\Gamma_q \left( (q-1) s\right) \leq I_q(s)$ by (\ref{gtauapp}).
Therefore if $I_q(s)$ is finite, then $\Gamma_q \left( (q-1) s\right)$ is finite, whence $\tau(q) \geq (q-1)s$, and finally $\tilde D_q \geq s$ by (\ref{dqtilde}).

\bibliography{population}

\end{document}